\documentclass[%
 aip,
 amsmath,amssymb,
 reprint,%
]{revtex4-1}

\usepackage{graphicx}

\usepackage{dcolumn}
\usepackage{bm}

\usepackage[utf8]{inputenc}
\usepackage[T1]{fontenc}
\usepackage{mathptmx}
\usepackage{etoolbox}

\usepackage{braket}
\usepackage{mathtools}

\usepackage{ulem}

\makeatletter
\def\@email#1#2{%
 \endgroup
 \patchcmd{\titleblock@produce}
  {\frontmatter@RRAPformat}
  {\frontmatter@RRAPformat{\produce@RRAP{*#1\href{mailto:#2}{#2}}}\frontmatter@RRAPformat}
  {}{}
}%
\makeatother
\begin{document}

\preprint{AIP/123-QED}

\title[Analytical Treatment of a Non-integrable Pendulum Based on Eigenvalue Problem of the Liouville Operator]
{Analytical Treatment of a Non-integrable Pendulum Based on Eigenvalue Problem of the Liouville Operator}
\author{Kosuke Asano}
 \email{szc04002@st.osakafu-u.ac.jp}
 \affiliation{Department of Physical Science, Osaka Prefecture University, 1-1 Gakuen-cho, Naka-ku, Sakai, Osaka 599-8531, Japan}
\author{Kenichi Noba}
 \email{noba@omu.ac.jp}
\affiliation{
Department of Physics, Osaka Metropolitan University, 1-1 Gakuen-cho, Naka-ku, Sakai, Osaka 599-8531, Japan
}
\author{Tomio Petrosky}
 \email{petrosky@austin.utexas.edu}
\affiliation{
Center for Complex Quantum Systems, University of Texas, Austin, Texas 6TX 78712, USA
}
 \altaffiliation[Also at]{
Institute of Industrial Science, University of Tokyo, Kashiwa, Chiba, 277-8574, Japan
}

\date{\today}

\begin{abstract}
We perform analytical and quantitative analysis of the motion of a non-integrable pendulum with two degrees of freedom, in which an integrable nonlinear  pendulum and a harmonic oscillator are weakly coupled through a non-integrable perturbative interaction, based on the eigenvalue problem of the Liouvillian that is the generator of time evolution in classical mechanics. The eigenfunctions belonging to the zero eigenvalue of the Liouvillian correspond to the invariants of the motion. The zero eigenvalue of the integrable unperturbed Liouvillian is infinitely degenerate at the resonance point. By applying a perturbation, level repulsion occurs between the eigenstates of the unperturbed system, and some of the degeneracy is lifted, resulting in a non-zero eigenvalue. In order to evaluate the frequency gap caused by the level repulsion,  we introduce an auxiliary operator called the collision operator which is well known in non-equilibrium statistical mechanics. We show that the dependence of the magnitude of frequency gap on the coupling constant can be quantitatively evaluated by simply finding the condition for the existence of the collision operator, without directly solving the eigenvalue problem.
\end{abstract}

\maketitle

\begin{quotation}
We analyze the motion of a non-integrable system using Liouvillian dynamics, which describes the time evolution of the phase space distribution function, rather than Hamiltonian dynamics, which focuses on trajectories. In classical mechanics, it is important to pay attention to invariants of motion in a system because their destruction by perturbations leads to chaotic behavior. In Liouvillian dynamics, where the Liouvillian is the generator of time evolution in the system, the invariants of motion correspond to the eigenstates associated with the zero eigenvalue of the Liouvillian. In our unperturbed system, the eigenstates with the zero eigenvalue are infinitely degenerate at resonance points. When a non-integrable perturbation is introduced, this degeneracy is lifted due to level repulsion at resonance, resulting in the destruction of these invariants of motion. In this paper, we analytically and quantitatively estimate the magnitude of the frequency gap in the eigenvalue spectrum of the Liouvillian caused by the level repulsion in a non-integrable pendulum with two degrees of freedom.
\end{quotation}

\section{Introduction}
\label{intro}

\begin{figure*}[t]
  \begin{center}
          \includegraphics[height=7cm,keepaspectratio]{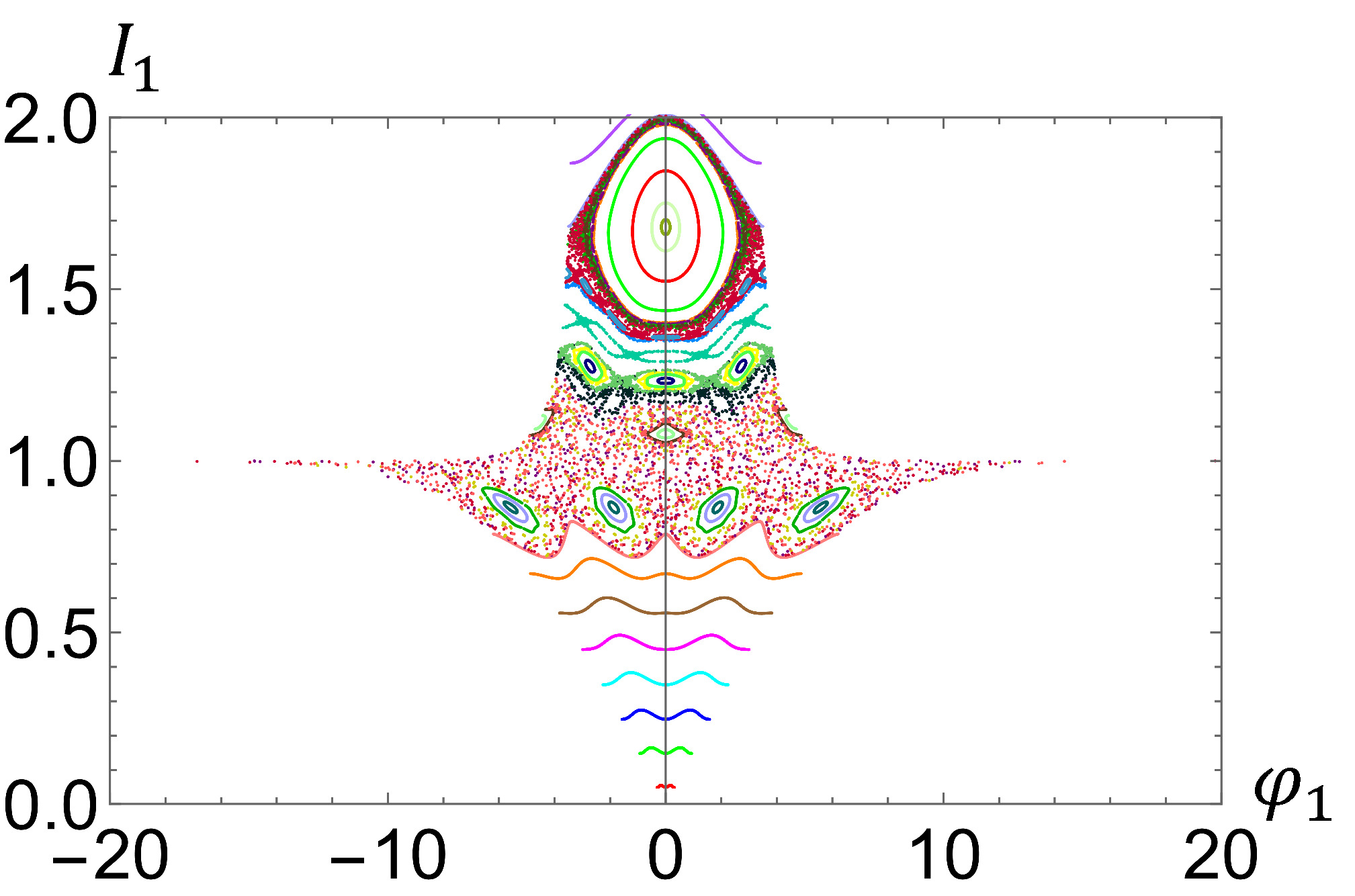}
\caption{Poincar\'e surface of section $(I_1, \varphi_1)$ of the non-integrable pendulum model with the Hamiltonian given by Eq.~(\ref{eq_ham1}). The variables of the axes, $\varphi_1$ and $I_1$, are explained in detail in the next section. A magnified version of the figure is shown in Fig. \ref{fig_poincareivarphi}.}
\label{fig_poincareivarphi_2}
  \end{center}
\end{figure*}

In this paper, we discuss a quantitative and analytical method for dealing with non-integrable few-degree-of-freedom systems in which chaos occurs, using the eigenvalue problem of the Liouville operator (also called the Liouvillian) that appears in the Liouville equation. This method is in contrast to the usual trajectory analysis based on Hamilton's equations of motion. As detailed in this paper, the method proposed in this paper is applicable to a wide range of non-integrable Hamiltonian systems with a few degrees of freedom. However, the analytical method introduced here incorporates many concepts that do not appear in conventional methods for handling non-integrable systems that focus on trajectories, such as collision operators that appear in nonequilibrium statistical mechanics and their nonlinear eigenvalue problems. In order to present the essence of these new concepts and methodologies as straightforward as possible, we present our method  by using a concrete example of non-integrable Hamiltonian system shown at Eq.\eqref{eq_ham1} that describes evolution of an actual nonlinear pendulum with two degrees of freedom \cite{Petrosky1984}, rather than discussing for the general Hamiltonian system.
\begin{eqnarray}
  {H}_t(P_{\Theta}, P_X, \Theta, X)&=&{H}_0(P_{\Theta}, P_X, \Theta, X) + \lambda {V}(\Theta, X),
    \label{eq_ham1}
\end{eqnarray}
with
\begin{eqnarray}
  {H}_0&=&{H}_1 + {H}_2,     
    \label{eq_ham10} \\
  {H}_1&=&\frac{P_{\Theta}^2}{2ml^2}+m\omega_0^2 l^2(1-\cos \Theta), 
    \label{eq_ham11} \\
  {H}_2&=&\frac{1}{2m}P_X^2+\frac{1}{2}m\omega_h^2 X^2,
    \label{eq_ham12} \\
  \lambda {V}&=&-\lambda mlX\omega_0^2 \cos \Theta ,
    \label{eq_ham13}
\end{eqnarray}
where $P_{\Theta}$ and $P_X$ are momenta which are canonical conjugate to the coordinates $\Theta$ and $X$, respectively. The parameter $m$ is the mass of the particle, $l$ is the length of the pendulum, $\omega_0=\sqrt{g/l}$ is the angular frequency of the pendulum with the gravitational acceleration $g$ when the amplitude is small, and $\omega_h$ is the natural frequency of the harmonic oscillator. The parameter $\lambda (\ge 0)$ is a dimensionless coupling constant. We consider the case $\lambda \ll 1$.
In the next section, we introduce new sets of canonical variables $(I_1,\varphi_1)$ and $(I_2,\varphi_2)$, which make the generalized coordinates $(\varphi_1,\varphi_2)$ cyclic variables in the unperturbed Hamiltonian $H_0$ (i.e., $H_0$ =$H_0(I_1,I_2)$). We show examples of the Poincar\'{e} surface of section in Figs. \ref{fig_poincareivarphi_2} and \ref{fig_poincareivarphi} which are obtained by numerical calculation using these new variables.

As is well known, in discussing problems in mechanics, it is important to identify invariants of motion that the system possesses. Furthermore, in analyzing non-integrable systems, it is important to quantitatively analyze how the invariants of motion in the unperturbed system are destroyed by the resonance effect due to the perturbation.
 However, from the definition of the term ``non-integrable'', it is easy to imagine that it is difficult to quantitatively treat the effects of the perturbation in a non-integrable system. The purpose of this paper is to show that two main results have been obtained regarding the quantitative treatment. The first purpose is to show how the difficult task of quantitatively analyzing this problem can be achieved not based on Hamilton's equations of motion that follows the evolution of trajectories, but instead by discussing the eigenvalue problem of the Liouvillian $L_H$, which is the generator of the time evolution that appears in the following Liouville equation that describes the time evolution of the state function of the system.

In general, the Liouville equation for a system with $\mathcal{N}$-degree of freedom is given by \cite{Goldstein2001}
\begin{eqnarray}
  i\frac{\partial}{\partial t}\rho(\bm{I}, \bm{\varphi},t)=L_H\rho(\bm{I}, \bm{\varphi},t),
    \label{eq_cle}
\end{eqnarray}
where $\bm{I}=(I_1, I_2, ..., I_\mathcal{N})$ is a generalized momentum vector with $\mathcal{N}$ components, $\bm{\varphi}=(\varphi_1, \varphi_2, ..., \varphi_\mathcal{N})$ is a generalized coordinate vector, and $\rho(\bm{I}, \bm{\varphi},t)$ is the state function defined in phase space. The Liouvillian $L_H$ is a linear operator and is represented by the Poisson Bracket with the Hamiltonian $H$ as
\begin{eqnarray}
  L_H \equiv i\{H,\hspace{7pt} \} 
       &=&i\left(\frac{\partial H}{\partial \bm{\varphi}} \cdot \frac{\partial}{\partial \bm{I}} - \frac{\partial H}{\partial \bm{I}} \cdot \frac{\partial}{\partial \bm{\varphi}}\right) .
    \label{eq_cl}
\end{eqnarray}
The imaginary unit $i$ in front of the Poisson Bracket in Eq.~(\ref{eq_cl}) has been introduced in order for the Liouvillian to become a Hermitian operator in the Hilbert space.  Therefore, the eigenvalue of the Liouvilian in the Hilbert space is a real number. The physical dimension of the Liouvilian is frequency.

By the way, when discussing motion, instead of discussing the time evolution of the state function $\rho(\bm{I},\bm{\varphi},t)$, it is also possible to consider the time evolution of a set of canonical variables $(\bm{I}(t),\bm{\varphi}(t))$ and discuss the time evolution of a physical quantity $A(\bm{I}(t), \bm{\varphi}(t))$ expressed as a function of those canonical variables. In that case, the equation for $A(\bm{I}, \bm{\varphi})$ is given by the following fundamental equation, which includes Hamilton's equations of motion as its special case,
\begin{equation}
  i\frac{d}{d t} A(\bm{I}, \bm{\varphi}) = -L_H A(\bm{I}, \bm{\varphi}) .
    \label{eq_ahameq}
\end{equation}
Note that the sign on the right side differs from that in Eq.~(\ref{eq_cle}).

These two different views in mechanics are complementary to each other. The view based on the state functions corresponds to the Schr\"{o}dinger picture of the probability density matrix in quantum mechanics, and the view based on the physical quantities corresponds to the Heisenberg picture of observables in quantum mechanics. In other words, there are two ways of analyzing mechanics in classical mechanics: the classical Schr\"{o}dinger picture and the classical Heisenberg picture. Our purpose in this paper is to discuss the relationship between the destruction of invariants of motion in non-integrable systems and resonance singularities based on the classical Schr\"{o}dinger picture.

As can be seen from Eqs. (\ref{eq_cle}) and (\ref{eq_ahameq}), an invariant of motion is the eigenstate belonging to the zero eigenvalue of the Liouvillian. When the total Hamiltonian $H$ is written as a sum of an unperturbed integrable Hamiltonian $H_0$ and a perturbation term representing the interactions among each degree of freedom, the eigenstates belonging to the zero eigenvalue of the unperturbed system are infinitely degenerate at the resonances, as will be shown later. Therefore, by analogy with the eigenvalue problem of the Hamiltonian in a quantum system where the eigenvalue is degenerate \cite{Kittel2005, Schiff1968, Peskin1995}, it is expected that the application of a perturbation may cause so-called level repulsion at the resonance point, and this degeneracy may be lifted \cite{Petrosky2011}. And due to this level repulsion, some of the eigenvalues associated with the invariant of motion in the unperturbed system become nonzero. As a result, it can be seen that the invariant of motion of the unperturbed system is destroyed by a non-integrable interaction. In this paper, we present a method to quantitatively evaluate the magnitude of the frequency gap caused by level repulsion around the zero eigenvalue of $L_H$.

However, the tricky thing here is that in the Liouvillian case, the eigenstates are not finitely degenerate, but infinitely degenerate. As a result, it is impossible to quantitatively evaluate the magnitude of the frequency gap around the resonance singularity in a non-integrable system of classical mechanics by solving algebraic equations of finite order, as is well known in quantum mechanics, for example. The second main purpose of this paper is to show how we solved this tricky point.

To solve this problem, we propose a method to solve the eigenvalue problem of an auxiliary operator called a collision operator (or effective Liouvillian) 
that is introduced by using a projection operator, rather than solving the Liouvillian eigenvalue problem directly. 
The collision operator has played a central role in exploring the basis of irreversibility in non-equilibrium statistical mechanics \cite{Prigogine1961, Petrosky1996,Balescu1963}.
In the usual method of directly solving the Liouvillian eigenvalue problem, 
all eigenvalues and eigenfunctions of the Liouvillian are found at once 
by appropriately rotating a huge matrix with off-diagonal elements obtained 
by displaying the Liouvillian using the base of the eigenvector of 
the unperturbed Liouvillian. 

However, our method introduces an operator that projects to the function space belonging to each eigenvalue of the unperturbed Liouvillian,
decomposes the eigenvalue problem into an eigenvalue problem for the projected subspace, and searches for the eigenvalues and eigenfunctions
of the original Liouvillian in the projected space. 
As a result, the function space that deals with the eigenvalue problem of the original Liouvillian is reduced to a subspace with fewer dimensions. However, on the other hand, 
although the original Liouvillian was a linear operator, 
the collision operator in the reduced subspace itself becomes 
a function of the eigenvalues of
the original Liouvillian, 
which brings about the complexity that the eigenvalue problem of the collision operator becomes a nonlinear eigenvalue problem with respect to the eigenvalue \cite{Petrosky1996}. When exploring the basis of various complex and interesting irreversible phenomena in non-equilibrium statistical mechanics, it all comes down to the problem of exploring the effects that emerge due to this nonlinearity.

Our problem here is not directly related to the problem of finding the basis of this irreversibility. Nevertheless, the second purpose of this paper is to show that by transforming the eigenvalue problem of the Liouvillian, which is a linear operator, into a nonlinear collision operator and utilizing its nonlinearity, we can quantitatively evaluate the coupling constant dependence of the magnitude of the frequency gap that occurs around the zero eigenvalue at the resonance without directly solving the eigenvalue problem.

To verify the validity of our results obtained by the theoretical analysis, we have  numerically solved Hamilton's equations of motion for the motion of the trajectory obtained from our Hamiltonian. And, taking the Fourier spectrum with respect to the frequency for the time evolution of generalized momentum on the trajectories, we  compare it with our theoretical result obtained from the consideration of eigenvalue problem of the Liouvillian. It was shown that the results obtained from the numerical calculations are in good agreement with the theoretical predictions.

The structure of this paper is as follows:

\noindent
In Section \ref{Nonintpendulum}, we introduce a new set of canonical variables that make the generalized coordinates $(\varphi_1,\varphi_2)$ cyclic variables in the unperturbed Hamiltonian $H_0$. 

\noindent
In Section \ref{liouville}, we introduce the eigenvalue problem of the Liouvillian for a general Hamiltonian consisting of an integrable system and a non-integrable interaction term. 

\noindent
In Section \ref{Degeneracy}, we discuss the relation between the resonance and the degeneracy of the eigenstate of the unperturbed Liouvillian for the case of the eigenvalue is zero. 

\noindent
In Section \ref{collisionoperator}, we formulate a nonlinear eigenvalue problem for a collision operator, which shares the same eigenvalues as the Liouvillian of the system.

\noindent
Section \ref{frequencygap} constitutes the primary focus of this paper; by discussing the convergence condition of the intermediate propagator within the collision operator, we estimate the size of the frequency gap in the eigenvalue of the Liouvillian.

\noindent
In Section \ref{Numerical}, we compare our theoretical estimation of the frequency gap with the one obtained by numerical calculation through the Fourier spectrum with respect to the frequency for the time evolution of generalized momentum on trajectories. 
 
\noindent
 In Section \ref{ResonanceAndHamiltonian}, we discuss the reason why the Hamiltonian is so special that it remains an invariant of motion for any conservative system despite the presence of resonance singularities from the perspective of the perturbation analysis.

\noindent
In Section \ref{concluding}, we provide a summary of this study and make some concluding remarks. 

\noindent
In Appendices, we prove some relations and also present several formulas that are used in the main part of the paper.

\section{The Non-integrable Pendulum
}
\label{Nonintpendulum}

In this section, we introduce the new canonical sets of variables $(I_1, \varphi_1)$ and $(I_2, \varphi_2)$, 
in which $\varphi_1, \varphi_2$ are cyclic coordinates for the unperturbed Hamiltonian.
We also introduce a dimensionless Hamiltonian $\tilde{H} \equiv H/ H_{\rm sx}$ and a dimensionless time $\tau \equiv \omega_0 t$. Here, the energy scale ${H}_{\rm sx}\equiv 2ml^2\omega_0^2$ represents the energy at the separatrix, which distinguishes rotational motion and librational motion of the unperturbed pendulum. By applying the canonical transformation shown below (see Eqs. (\ref{eq_thetacano}) to (\ref{eq_pxcano}) and Appendix \ref{section_B}), we obtain the dimensionless Hamiltonian,
\begin{eqnarray}
 \tilde{H}(I_1, I_2, \varphi_1,\varphi_2)
  =\tilde{H}_0(I_1, I_2) + \lambda \tilde{V}(I_1, I_2, \varphi_1,\varphi_2)
    \label{eq_ham2}
\end{eqnarray}
with
\begin{eqnarray}
  \tilde{H}_0 &=&  \tilde{H}_1 +  \tilde{H}_2,     
    \label{eq_ham20} \\
  \tilde{H}_1&=&{H}_1/{H}_{\rm sx}=I_1^2, 
    \label{eq_ham21} \\
   \tilde{H}_2&=&{H}_2/{H}_{\rm sx}=I_2 / \kappa, 
    \label{eq_ham22} \\
  \lambda  \tilde{V}&=&\lambda {V}/{H}_{\rm sx} 
    = \lambda \sqrt{\kappa I_2} f(I_1, \varphi_1) \cos \varphi_2, 
      \label{eq_ham23}
\end{eqnarray}
where the unperturbed Hamiltonian $ \tilde{H}_0$ depends only on the generalized momenta $I_1$ and $I_2$.
Here, 
\begin{eqnarray}
  \kappa&\equiv&\omega_0/\omega_h ,
    \label{eq_chi}
\end{eqnarray}
(see Eqs. (\ref{eq_ham11}) and (\ref{eq_ham12})) and
\begin{eqnarray}
  f(I_1, \varphi_1)&\equiv&
    \begin{cases}
      2{\rm cn}^2 \left( \varphi_1 /2, c \right) -1 & c \le 1 \\
      2{\rm dn}^2 \left( \frac{c}{2}(\varphi_1-\varphi_{1,0}), 1/c \right) -1 & c >1,
    \end{cases} \label{eq_ham23p} 
\end{eqnarray}
where cn and dn are the Jacobi elliptic functions with the modulus $c$ given  by
\begin{eqnarray}
  c&\equiv&1/I_1=1/\sqrt{\tilde{H}_1}  .
    \label{eq_cy1}
\end{eqnarray}
The pendulum rotates when $I_1 \ge 1$ (i.e., $c \le 1$), and oscillates when $I_1 < 1$ (i.e., $c > 1$). The constant $\varphi_{1,0}$ in Eq. (\ref{eq_ham23p}) is defined by
\begin{eqnarray}
  \varphi_{1,0}\equiv
    \begin{cases}
      0 & P_{\Theta}\ge 0 \\
      \pi/\Delta_1(I_1) & P_{\Theta}<0 ,
    \end{cases} \label{eq_alpha0}
\end{eqnarray}
which is introduced to ensure one complete cycle of the librational motion of the pendulum (see Appendix \ref{section_B}). 
Here, the quantity $\Delta_1(I_1)$ is defined as
\begin{eqnarray}
  \Delta_1(I_1)&\equiv&
    \begin{cases}
      \frac{\pi}{2K(1/I_1)} & c \le 1 \\
      \frac{\pi }{4I_1K(I_1)} & c>1,
    \end{cases} \label{eq_kappa1}
\end{eqnarray}
where $K(c)$ or $K(1/c)$ is the complete elliptic integral of the first kind.

The new variables $(I_1, I_2, \varphi_1, \varphi_2)$ are related to the old variables $(P_{\Theta},P_X, \Theta, X)$ through the canonical transformations shown in Appendix \ref{section_B}, as
\begin{eqnarray}
  \sin \frac{\Theta}{2}&=&
    \begin{cases}
      \pm {\rm sn} (\frac{1}{2} \varphi_1, c) & c \le 1 \\
      \pm \frac{1}{c}{\rm sn} \left( \frac{c}{2} (\varphi_1-\varphi_{1,0}), 1/c \right) & c>1,
    \end{cases}
    \label{eq_thetacano} \\
  P_{\Theta}&=&
    \begin{cases}
      \pm \frac{{H}_{\rm sx}}{\omega_0 c} {\rm dn} (\frac{1}{2} \varphi_1, c) & c \le 1 \\
      \pm \frac{{H}_{\rm sx}}{\omega_0 c} {\rm cn} \left( \frac{c}{2} (\varphi_1-\varphi_{1,0}), 1/c \right) & c>1,
    \end{cases}
    \label{eq_pthetacano} \\
  X&=&\sqrt{\frac{2 {H}_{\rm sx} I_2}{m \kappa \omega_h^2}} \cos \varphi_2 ,
    \label{eq_xcano} \\
  P_X&=&-\sqrt{\frac{2m {H}_{\rm sx} I_2}{\kappa}} \sin \varphi_2 ,
    \label{eq_pxcano} 
\end{eqnarray}
where ${\rm sn}$ is also a Jacobi elliptic function. In order to avoid too heavy notations, we will drop the tilde notation on the dimensionless Hamiltonian $\tilde{H}$ and the dimensionless interaction $\lambda \tilde{V}$, and from now on we will use the same notation such as $H$ and $\lambda V$ as before instead of $\tilde{H}$ and $\lambda \tilde{V}$, respectively. 

By using the Fourier cosine series expansion of ${\rm sn}^2(u, c)$,
\begin{eqnarray}
  {\rm sn}^2(u, c) 
  &=& \frac{1}{c^2}\left( 1-\frac{E(c)}{K(c)} \right) \nonumber \\
  & & -\frac{\pi}{c^2 K(c)} \sum_{n=1}^{\infty} \frac{\pi n/K(c)}{\sinh (\pi n K'(c)/K(c))} \cos \left( \frac{n \pi u}{K(c)} \right) , \nonumber \\
    \label{eq_sn2}
\end{eqnarray}
with ${\rm cn}^2(u, c)=1-{\rm sn}^2(u, c)$ and ${\rm dn}^2(u, c)=1-c^2{\rm sn}^2(u, c)$, the Fourier series expansion of the interaction in Eq.~(\ref{eq_ham23}) is given as 
\begin{eqnarray}
  \lambda V(I_1, I_2, \varphi_1, \varphi_2)=\lambda \sum_{n_1} \sum_{n_2=\pm 1} V_{n_1, n_2}(I_1, I_2) e^{i[n_1\Delta_1(I_1) \varphi_1 + n_2 \varphi_2]} , \nonumber \\ \label{eq_vfourierexpansion00}
\end{eqnarray}
where
\begin{eqnarray}
  & &\hspace{-10pt} V_{n_1, \pm 1}(I_1, I_2) \nonumber \\ 
  &=&
    \begin{cases}
      -\sqrt{\kappa I_2} \cfrac{1}{2} \left\{1-2 I_1^2 \left(1-\frac{E(1/I_1)}{K(1/I_1)}\right)\right\} & c\le1, n_1=0 \\
      -\sqrt{\kappa I_2} \cfrac{2 n_1 I_1^2 \Delta_1^2(I_1)}{ \sinh (2n_1\Delta_1(I_1)K'(1/I_1))} & c\le1, n_1\ne 0 \\
      -\sqrt{\kappa I_2} \cfrac{1}{2} \left\{1-2\left(1-\frac{E(I_1)}{K(I_1)}\right)\right\} & c>1, n_1=0 \\
      -\sqrt{\kappa I_2} \cfrac{4 n_1 I_1^2 \Delta_1^2(I_1)}{ \sinh (2n_1 I_1\Delta_1(I_1)K'(I_1))} & c>1, n_1 {\rm \hspace{3pt}is \hspace{3pt}even} \\
      0 & c>1, n_1 {\rm \hspace{3pt}is \hspace{3pt}odd} 
    \end{cases} \nonumber \\  \label{eq_vfourierexpansion}
\end{eqnarray}
with
\begin{eqnarray}
  K'(c) \equiv K(\sqrt{1-c^2}),
    \label{eq_ellipticKp}
\end{eqnarray}
and $E(c)$ is a complete elliptic integral of the second kind.

Although the interaction in Eq. (\ref{eq_ham13}) has a relatively simple form, the fact that this system is non-integrable can be seen from the appearance of chaotic regions in the Poincar\'e surfaces of section shown in Fig. \ref{fig_poincareivarphi_2}. This fact can also be seen from the form of the new Hamiltonian in which the unperturbed terms are represented only by the momenta, and the interaction in Eqs.~(\ref{eq_vfourierexpansion00}) and (\ref{eq_vfourierexpansion}) is written as the sum of infinitely many trigonometric functions with argument $(n_1 \Delta_1 \varphi_1 \pm \varphi_2)$. As is well known, if the trigonometric part consists of only one term, the system is integrable. In such a case, the Hamiltonian is called a single-resonance Hamiltonian\cite{WalkerFord1969}. On the other hand, if the trigonometric part is composed of a sum of trigonometric functions for arguments of $n_1$ with various values, the Hamiltonian is called a multiple-resonance Hamiltonian. It is known that a multiple-resonance Hamiltonian is generically a non-integrable system \cite{WalkerFord1969}.

\begin{figure*}[t]
  \begin{center}
          \includegraphics[height=7cm,keepaspectratio]{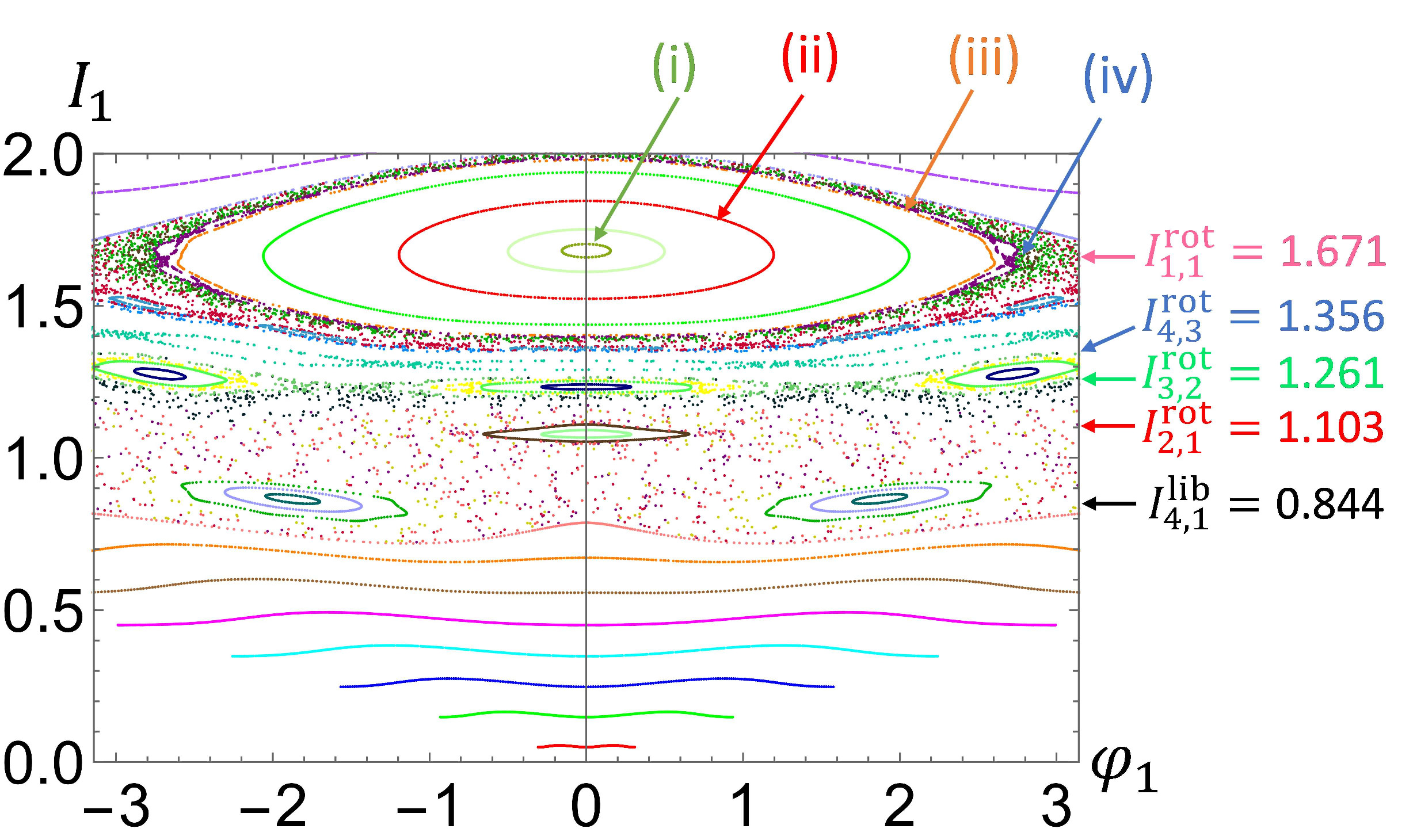}
\caption{ This is a magnified figure of the Poincar\'e surface of section $(I_1, \varphi_1)$ in Fig. \ref{fig_poincareivarphi_2} for our non-integrable pendulum model, where the dimensionless Hamiltonian $H=18.222$ in case of $\lambda=1/10$, $\kappa=1/3$ and $d\varphi_2 / d\tau>0$. The intersections of the surface and the trajectory at $\varphi_2=2\pi \times \tilde{m}$, where $\tilde{m}$ is an integer, are plotted. The $1/1, 4/3, 3/2$ and $2/1$ resonance points for the rotation of the pendulum are at $I_1=I^{\rm rot}_{1,1}, I^{\rm rot}_{4,3}, I^{\rm rot}_{3,2}$ and $I^{\rm rot}_{2,1}$, respectively, while the $4/1$ resonance point for the libration of the pendulum is at  $I_1 = I^{\rm lib}_{4,1}$. Trajectories corresponding to (i), (ii) and (iii) exhibit almost regular motions with the intersections on the surface of section aligning on one-dimensional curved lines around the $1/1$ primary resonance, while trajectory corresponding to (iv) represents chaotic motion with intersections scattered as points.
}
\label{fig_poincareivarphi}
  \end{center}
\end{figure*}
\begin{figure*}[t]
  \begin{center}
    \includegraphics[height=7cm,keepaspectratio]{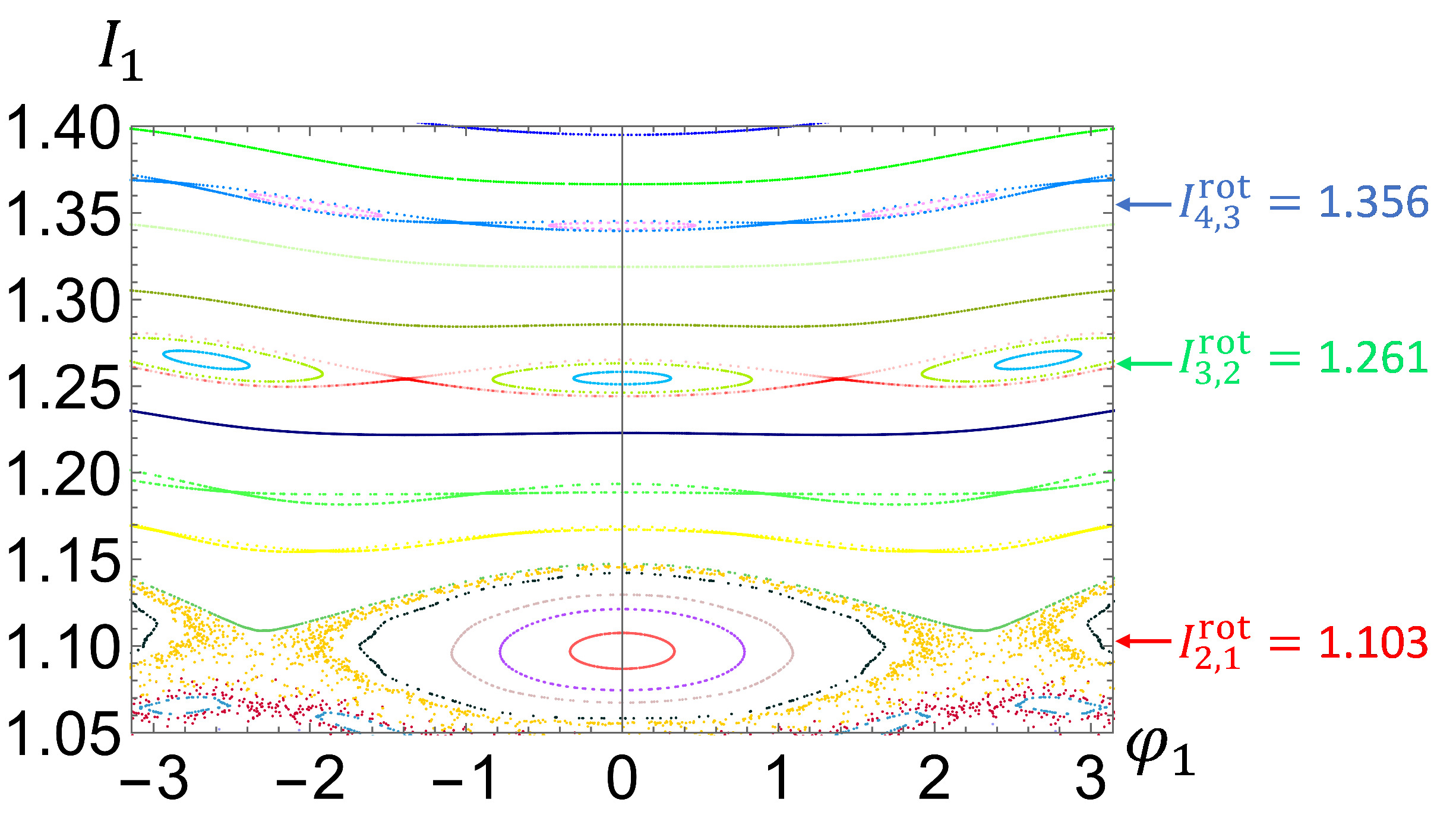}
 \caption{
This is a part of the Poincar\'e surface of section $(I_1, \varphi_1)$ for the non-integrable pendulum model for a smaller value of $\lambda=1/40$ as compared to Fig.~\ref{fig_poincareivarphi_2} with the same values $H=18.222$, $\kappa=1/3$ and $d\varphi_2 / d\tau>0$. It can be observed that the smaller $\lambda$ is and the larger $N$ in $I_{M,N}$ is, the smaller the resonance effect is.
   }
          \label{fig_poincareivarphi40}
  \end{center}
\end{figure*}

We show the Poincar\'e surface of section $(\varphi_1, I_1)$ for this non-integrable pendulum system in Fig.~\ref{fig_poincareivarphi}. This figure is a magnification of Fig. \ref{fig_poincareivarphi_2}. At the separatrix, where
\begin{eqnarray}
 I_1 =1 ,
 \label{separatrix}
\end{eqnarray}
the period of $\varphi_1$ becomes infinity. This is in contrast to the action-angle variable. Indeed, if we  use the action-angle variable for the unperturbed Hamiltonian of the non-integrable pendulum, we would have a discontinuity in the the Poincar\'e surface of section at the separatrix, since one cannot map the infinitely large period to $2\pi$ period of the angle variable. On the other hand, the use of our new variable $\varphi_1$ can avoid this difficulty because the infinite period of  $\varphi_1$ is consistent with the period of the separatrix (see Appendices \ref{section_B} and \ref{section_a}).

We show  the case of the value of the dimensionless Hamiltonian $H=18.222$ for  $\lambda=1/10$, $\omega_0=10$, and $\omega_h=30$ ($\kappa=1/3$) in Fig. \ref{fig_poincareivarphi}. In this figure we have chosen $\kappa=1/3$ in such away that the frequency of harmonic oscillator has the same order of magnitude as the frequency of the non-linear pendulum when the amplitude of the pendulum is small. In this figure, one can see a structure with several islands that is a typical structure in the Poincar\'e surfaces for non-integrable systems. 
Each island is distinguished by the type of resonance determined by the integer ratio between the frequencies associated with
 each degree of freedom. As will be explained later in detail at Eq. \eqref{resonanceCondition}, our two-degree-of-freedom system is characterized by 
a pair of coprime integers $M$ and $N$ such that $\omega_2/\omega_1=M/N$, where $\omega_1$ and $\omega_2$ are the frequencies of each degree of freedom. Therefore, we call this resonance the $M/N$ resonance. Furthermore, in nonlinear systems such as those considered here, the frequency generally depends on the momentum, so the system is also characterized by the value of momentum that satisfies the ratio of the frequencies under the condition of $M/N$ resonance. Here, the symbol $I_{M,N}$ indicates the value of the generalized momentum that satisfies the condition of this $M/N$ resonance.

The different colors of the points plotted on the surface of section 
in Fig. \ref{fig_poincareivarphi} 
correspond to different initial conditions. 
When $I_1\lesssim 0.7$, the sequence of the points lies on one-dimensional curved lines. 
Therefore, the non-integrable pendulum for $I_1\lesssim 0.7$ exhibits almost regular motion
in our resolution. 
For $I_1\gtrsim 0.7$, the chaotic motions of the non-integrable pendulum are visible as scattered points 
on the surface of section \cite{Henon1964, Chirikov1979}.

In Fig.~\ref{fig_poincareivarphi40}, we show a part of the Poincar\'e surface of section for our system with a smaller value of  $\lambda=1/40$, as compared to Fig.~\ref{fig_poincareivarphi}, under the same conditions $H=18.222$, $\kappa=1/3$ and $d\varphi_2 / d\tau>0$.  It can be seen that as $\lambda$ decreases and $N$ in $I_{M,N}$ increases,  the island structure at the resonance becomes smaller, i.e., the resonance effect diminishes. Hence, we will refer to the case $N=1$ as the primary resonance, the case $N=2$ as the secondary resonance, and so on. Indeed, we will theoretically show in Section \ref{frequencygap}
that the larger the value of $N$, the smaller the magnitude of the frequency gap of the eigenvalue in our non-integrable pendulum.

\section{Eigenvalue Problem of the Liouvillian}
\label{liouville}

In this section, we analyze the eigenvalue problem of the Liouvillian. We first start with a class of more general nonlinear Hamiltonian systems with $\mathcal{N}$ degrees of freedom. The Hamiltonian of this class of systems is expressed as the sum of an integrable system $H_0$ and an interaction term with a dimensionless non-vanishing coupling constant $\lambda V$ that makes the system non-integrable, and can be written as follows:
\begin{eqnarray}
  H(\bm{I}, \bm{\varphi})&=&H_0(\bm{I})+\lambda V(\bm{I}, \bm{\varphi}) .
    \label{eq_h}
\end{eqnarray}
Corresponding to this sum of Hamiltonians, the Liouvillian can also be written as a sum of two parts:
\begin{eqnarray}
  L_H&=&L_0+\lambda L_V ,
    \label{eq_sepl}
\end{eqnarray}
where each term is expressed using the Poisson bracket: 
\begin{eqnarray}
  L_0&=&i\{H_0,\hspace{7pt} \},  \\
  L_V&=&i\{V,\hspace{7pt} \} .
    \label{eq_jv}
\end{eqnarray}

In this class of systems, we assume that the unperturbed Hamiltonian is separated for each degree of freedom,
\begin{eqnarray}
  H_0(\bm{I})&=&\sum_{j=1}^\mathcal{N} H_j(I_j) .
    \label{eq_h0}
\end{eqnarray}
In this case, the unperturbed Liouvillian is given by
\begin{eqnarray}
  L_0 = -i\bm{\nu} \cdot \frac{\partial}{\partial \bm{\varphi}}
      =-i \sum_{j=1}^\mathcal{N} \nu_j(I_j) \frac{\partial}{\partial \varphi_j},
    \label{eq_j0}
\end{eqnarray}
where
\begin{eqnarray}
  \nu_j(I_j)&\equiv&\frac{\partial H_j(I_j)}{\partial I_j} ,
    \label{eq_omega}
\end{eqnarray}
and  $\bm{\nu}=(\nu_1(I_1), \nu_2(I_2), ..., \nu_\mathcal{N}(I_\mathcal{N}))$. Since our system is nonlinear, some of the components $\nu_j(I_j)$ depend on $I_j$.

Before discussing the eigenvalue problem for the total Liouvillian $L_H$, we first present the solution to the eigenvalue problem of the unperturbed Liouvillian $L_0$, 
\begin{eqnarray}
  L_0  \phi_0^{(\bm{n})} (\bm{I}, \bm{\varphi})=w^{(\bm{n})} (\bm{I}) \phi_0^{(\bm{n})} (\bm{I}, \bm{\varphi}) .
    \label{eq_evp00}
\end{eqnarray}
The eigenvalue and eigenfunction are given by
\begin{eqnarray}
  w^{(\bm{n})} (\bm{I})
    &=& \sum_{j=1}^\mathcal{N} n_j \Delta_j(I_j) \nu_j(I_j)
    \equiv \bm{k}(\bm{n}, \bm{I}) \cdot \bm{\nu} ,
    \label{eq_ev0} \\
  \phi_0^{(\bm{n})} (\bm{I}, \bm{\varphi})
    &=&\sqrt{\frac{\prod_{j=1}^\mathcal{N} \Delta_j(I_j)}{(2\pi)^{\mathcal{N}}}} e^{i \sum_{j=1}^\mathcal{N} n_j \Delta_j(I_j) \varphi_j} \nonumber \\
    &=&\sqrt{\frac{\prod_{j=1}^\mathcal{N} \Delta_j(I_j)}{(2\pi)^{\mathcal{N}}}} e^{i \bm{k}(\bm{n}, \bm{I}) \cdot \bm{\varphi}} ,
      \label{eq_ef0}
\end{eqnarray}
where $\bm{n}=(n_1, n_2, \cdots, n_\mathcal{N})$ and $n_j$ is an integer ranging from  $-\infty $ to $+\infty$, and $\bm{k}(\bm{n}, \bm{I})$ is a vector whose component is $k_j(n_j, I_j)\equiv n_j \Delta_j(I_j)$. Here, $\Delta_j(I_j)$ represents a discreteness of the spectrum of $k_j(n_j, I_j)$, which is determined by the period $2\pi/\Delta_j(I_j)$ of the eigenfunction (\ref{eq_ef0}) with respect to the variable $\varphi_j$. 
The factor in front of  Eq. (\ref{eq_ef0}) is the normalization constant.

We consider the case where $\Delta_j(I_j)$ is finite, so the eigenvalues of the unperturbed Liouvillian are discrete for each value of the generalized  momenta $I_j$ \cite{Petrosky2004, ChynBiu2004}. If we introduce the variables $\theta_j(I_j, \varphi_j) \equiv \Delta_j(I_j) \varphi_j$, we have the angle variable $\theta_j$ with the period $2\pi$, and the angular frequency $\omega_j(I_j) \equiv \Delta_j(I_j) \nu_j(I_j)$, respectively. Note that if $\Delta_j(I_j)$ depends on $I_j$, $I_j$ is {\it not} the action variable that is canonical conjugate to the angle variable. However, using these angular frequencies and angle variables, we can write the eigenvalue and the eigenfunction as
\begin{eqnarray}
  w^{(\bm{n})} (\bm{I})
    &=& \bm{n} \cdot \bm{\omega} ,
      \label{eq_ev00} \\
  \phi_0^{(\bm{n})} (\bm{I}, \bm{\varphi})
    &=& \sqrt{\frac{\prod_{j=1}^\mathcal{N} \Delta_j(I_j)}{(2\pi)^{\mathcal{N}}}} e^{i \bm{n} \cdot \bm{\theta} },
      \label{eq_ef00}
\end{eqnarray}
where $\bm{\omega}=(\omega_1(I_1), \omega_2(I_2), \cdots, \omega_\mathcal{N}(I_\mathcal{N}))$, and $\bm{\theta}=(\theta_1(I_1, \varphi_1), \theta_2(I_2, \varphi_2), ..., \theta_\mathcal{N}(I_\mathcal{N}, \varphi_\mathcal{N}))$. Action variable $\bm{J}$ that is canonical conjugate to the angle variable  $\bm{\theta}$  will be introduced in Appendix \ref{section_a}. The following discussion uses both the variables $\bm{\varphi}$ and $\bm{\theta}$ depending on the context of the discussion.

Since the unperturbed Liouvillian is a Hermitian operator, the eigenfunctions of the unperturbed Liouvillian form an orthonormal and complete set, i.e., 
\begin{eqnarray}
  \int_{-\pi/\bm{\Delta} (\bm{I})}^{\pi/\bm{\Delta} (\bm{I})} d\bm{\varphi} \phi_0^{{(\bm{n})}*} (\bm{I}, \bm{\varphi}) \phi_0^{(\bm{n'})} (\bm{I}, \bm{\varphi})
    &=&\delta_{n_1,n_1'} \delta_{n_2,n_2'} \cdots \delta_{n_\mathcal{N},n_\mathcal{N}'} \nonumber \\
    &\equiv& \delta_{\bm{n},\bm{n'}},
    \label{eq_orthonomal0}
\end{eqnarray}
\begin{eqnarray}
  & &\hspace{-10pt} \sum_{\bm{n}} \phi_0^{{(\bm{n})}*} (\bm{I}, \bm{\varphi}) \phi_0^{(\bm{n})} (\bm{I}, \bm{\varphi'}) \nonumber \\
    &=&\delta (\varphi_1-\varphi'_1) \delta (\varphi_2-\varphi'_2) \cdots \delta (\varphi_\mathcal{N}-\varphi'_\mathcal{N}) \equiv \delta (\bm{\varphi}-\bm{\varphi'}) , \nonumber \\
    \label{eq_completenessrelation0}
\end{eqnarray}
where
\begin{eqnarray}
  & & \hspace{-10pt} \int_{-\pi/\bm{\Delta} (\bm{I})}^{\pi/\bm{\Delta} (\bm{I})} d\bm{\varphi} \nonumber \\
   &=& 
   \int_{-\pi/\Delta_1(I_1)}^{+\pi/\Delta_1(I_1)}d\varphi_1 
   \int_{-\pi/\Delta_1(I_2)}^{+\pi/\Delta_1(I_2)}d\varphi_2 
   \, \cdots \, 
   \int_{-\pi/\Delta_\mathcal{N}(I_\mathcal{N})}^{+\pi/\Delta_\mathcal{N}(I_\mathcal{N})}d\varphi_\mathcal{N},  \nonumber \\
    \label{eq_intdphi}
\end{eqnarray}
the notation * stands for the complex conjugate, and we have used the following abbreviated notation,
\begin{eqnarray}
  \sum_{\bm{n}} \equiv \sum_{n_1=-\infty}^{\infty} \sum_{n_2=-\infty}^{\infty} \cdots \sum_{n_\mathcal{N}=-\infty}^{\infty}.
    \label{eq_nreduction}
\end{eqnarray}

 Let us note that due to the completeness relation (\ref{eq_completenessrelation0}),  any function of $\bm{\varphi}$ can be expanded in terms of the eigenfunctions (\ref{eq_ef0}) or (\ref{eq_ef00}). Due to the form given in Eq. (\ref{eq_ef00}), this expansion is simply a Fourier transformation with respect to the angle variable $\bm{\theta}$. Hence, this completeness relation is confined within the space spanned solely by the angle variable $\bm{\theta}$, or the variable $\bm{\varphi}$, for any value of the generalized momentum $\bm{I}$. In this case, it is possible to treat the angle variable $\bm{\theta}$, or the variable $\bm{\varphi}$, separately from the generalized momentum $\bm{I}$, and to consider the generalized momentum as a parameter instead of a variable as explained in the footnote of p23 in Ref.~\cite{Prigogine1961}. Hereafter, we will treat the generalized momentum as a parameter. However, it should be noted that in this treatment, the matrix element of the Liouvillian obtained using the $\bm{n}$ representation is not  simply a  number, but generally a derivative operator with respect to the generalized momentum (see Eqs. (\ref{eq_matrixelementprime1}) and (\ref{eq_matrixelementactionangle})).

With this in mind, we introduce the bra-ket notation commonly used in quantum mechanics for the variable $\bm{n}$ conjugate  to the variable $\bm{\varphi}$.
Then, in terms of the bra-ket notation, the eigenvalue equation for the unperturbed Liouvillian is written as
\begin{eqnarray}
L_0 |\bm{n} ; \bm{I}\rangle = w^{(\bm{n})}(\bm{I}) |\bm{n}; \bm{I}\rangle ,
    \label{eq_evp000} 
\end{eqnarray}
and the eigenfunction is defined as
\begin{eqnarray}
  \phi_0^{(\bm{n})} (\bm{I}, \bm{\varphi}) \equiv \langle \bm{\varphi} | \bm{n} ; \bm{I}\rangle .
    \label{eq_ef0000}
\end{eqnarray}
From now on, for brevity, we will abbreviate $|\bm{n}; \bm{I}\rangle$ as $|\bm{n}\rangle$. Thus we have
\begin{eqnarray}
\langle \bm{n} | \bm{n'} \rangle = \delta_{\bm{n},\bm{n'}},
    \label{eq_orthonomal0braket} \\
\sum_{\bm{n}} | \bm{n} \rangle \langle \bm{n} | = \hat{I}_{\bm{\varphi}},
    \label{eq_completenessrelation0braket}
\end{eqnarray}
where $\hat{I}_{\bm{\varphi}}$ is the unit operator in the vector space associated with the generalized coordinate $\bm{\varphi}$.

Let us now turn to our main problem of the eigenvalue problem of the perturbed Liouvillian given by the following equation,
\begin{eqnarray}
  L_H \Phi_{\alpha}^{(\bm{n})} (\bm{I}, \bm{\varphi}) = \Omega_{\alpha}^{(\bm{n})}(\bm{I}) \Phi_{\alpha}^{(\bm{n})} (\bm{I}, \bm{\varphi}) ,
    \label{eq_evp1}
\end{eqnarray}
where $\alpha$ is the index associated with the degeneracy in the unperturbed system.
In the bra-ket notation, we have
\begin{eqnarray}
  L_H | \Phi_{\alpha}^{(\bm{n})}; \bm{I} \rangle=\Omega_{\alpha}^{(\bm{n})}(\bm{I}) | \Phi_{\alpha}^{(\bm{n})}; \bm{I} \rangle ,
    \label{eq_evp}
\end{eqnarray}
where the eigenfunction is defined as
\begin{eqnarray}
  \Phi_{\alpha}^{(\bm{n})} (\bm{I}, \bm{\varphi}) \equiv \langle \bm{\varphi} |  \Phi_{\alpha}^{(\bm{n})}; \bm{I} \rangle .
    \label{eq_ef000}
\end{eqnarray}
From now on, for brevity, we will abbreviate $| \Phi_{\alpha}^{(\bm{n})} \rangle = |\Phi_{\alpha}^{(\bm{n})}; \bm{I} \rangle$. The orthonormality and completeness relations of the perturbed system are written as
\begin{eqnarray}
  \langle\Phi_{\alpha}^{(\bm{n})} | \Phi_{\alpha'}^{(\bm{n'})} \rangle  = \delta_{\bm{n},\bm{n'}} \delta_{\alpha, \alpha'} ,
    \label{eq_orthonomal1braket} \\
  \sum_{\bm{n}} \sum_{\alpha} | \Phi_{\alpha}^{(\bm{n})} \rangle \langle \Phi_{\alpha}^{(\bm{n})} | = \hat{I}_{\bm{\varphi}} .
    \label{eq_completenessrelation1braket}
\end{eqnarray}

Let us expand the interaction $\lambda V(\bm{I}, \bm{\varphi})$ by the unperturbed eigenstates of $L_0$ as
\begin{eqnarray}
  \lambda V(\bm{I}, \bm{\varphi}) 
   = \lambda {\sum_{\bm{n}}}^{'} V_{\bm{n}}(\bm{I}) e^{i \bm{k}(\bm{n},\bm{I}) \cdot \bm{\varphi}}
   = \lambda {\sum_{\bm{n}}}^{'} V_{\bm{n}}(\bm{I}) e^{i \bm{n} \cdot \bm{\theta}}, 
    \label{eq_jamvp}
\end{eqnarray}
where the prime sign in ${\sum}^{'}$ stands for excluding the component with $\bm{n}=\bm{0}$ by assuming the component $\lambda V_{\bm{0}}(\bm{I}) =0$. If this component is not zero,
 we can redefine the unperturbed Hamiltonian by incorporating this $\bm{\varphi}$-independent component $\lambda V_{\bm{0}}(\bm{I})$ into the unperturbed Hamiltonian $H_0$. 
 Note that this is nothing but the Fourier expansion in terms of the angle variable $\bm{\theta}$.

The matrix element of the perturbation term of the Liouvillian is given by
\begin{eqnarray}
  \langle\bm{n} | L_V | \bm{n'}\rangle
    = \int_{-\pi/\bm{\Delta} (\bm{I})}^{\pi/\bm{\Delta} (\bm{I})} d\bm{\varphi} \phi_0^{{(\bm{n})}*} (\bm{I}, \bm{\varphi}) L_V \phi_0^{(\bm{n'})} (\bm{I}, \bm{\varphi}).
    \label{eq_matrixelementprime}
\end{eqnarray}
From Eq. (\ref{eq_jamvp}) and 
\begin{eqnarray}
 \int_{-\pi/\Delta_j}^{\pi/\Delta_j} d \varphi_j    e^{i (n_j-n_j') \Delta_j \varphi_j} = \frac{2\pi}{\Delta_j} \delta_{n_j,n_j'},
    \label{eq_en-np}
\end{eqnarray}
we 
obtain
\begin{eqnarray}
  \langle\bm{n} | L_V | \bm{n'} \rangle
    &=& - \left( \frac{1}{2} V_{\bm{n}-\bm{n'}}(\bm{I}) (\bm{n}-\bm{n'})\cdot \bm{\Delta}'(\bm{I}) \right. \nonumber \\
    & &\left. + V_{\bm{n}-\bm{n'}}(\bm{I}) \left( \bm{k}(\bm{n}, \bm{I})-\bm{k}(\bm{n'}, \bm{I}) \right) \cdot \frac{\partial}{\partial \bm{I}} \right. \nonumber \\
    & &\left. - \bm{k}(\bm{n'}, \bm{I}) \cdot \frac{\partial V_{\bm{n}-\bm{n'}}(\bm{I})}{\partial \bm{I}} \right),
    \label{eq_matrixelementprime1}
\end{eqnarray}
where  $\bm{\Delta}' (\bm{I})\equiv (\partial \Delta_1(I_1) / \partial I_1, \partial \Delta_2(I_2) / \partial I_2, \cdots, \partial \Delta_1(I_\mathcal{N}) / \partial I_\mathcal{N} )$. Note that the matrix element in this representation still consists of the differential operator with respect to the parameter $\bm{I}$ as mentioned above.

Here, we have shown the matrix element of the perturbed Liouvillian in the $(\bm{I},\bm{n})$ representation based on the $(\bm{I},\bm{\varphi})$ variable. We can also represent this matrix element in the $(\bm{J},\bm{n})$ representation based on the action-angle $(\bm{J},\bm{\theta})$ variable. In this case, the matrix elements have a seemingly simpler structure, as shown in (\ref{eq_matrixelementactionangle}) in Appendix \ref{section_a}, compared to Eq.~(\ref{eq_matrixelementprime1}).
Nevertheless, our variables $(\bm{I}, \bm{\varphi})$ have an advantage compared to the action-angle variables as explained in \eqref{separatrix}
 and at the end of Appendix \ref{section_a}. As shown in Appendix \ref{section_D}, which of these representations is more convenient to use depends on the context in which the formulas are used.

\section{Infinite degeneracy of the unperturbed eigenvalue} 
\label{Degeneracy}

As mentioned in the introduction, we are interested in how the invariant of motion in the unperturbed system is destroyed 
by imposing the perturbation.
 From the perspective of the eigenvalue problem of the Liouvillian, we can understand this as a problem of how the eigenfunctions associated with the zero eigenvalue in the unperturbed system are affected 
by imposing the perturbation.
According to Eq. (\ref{eq_ev00}), the eigenvalue of the unperturbed Liouvillian is 
zero in the case of $\bm{n}=\bm{0}$. 
The corresponding eigenfunctions do not depend on $\bm{\varphi}$, i.e., they depend only on the generalized momentum $\bm{I}$, which is the invariant of motion in the unperturbed system as seen from the canonical equations of motion.

In addition, let us note that if the resonance condition $\bm{n} \cdot \bm{\omega}=0$ is satisfied for $\bm{n}\not= \bm{0}$, Eq. (\ref{eq_ev00}) also shows that the unperturbed eigenvalue is zero.
For example, when the unperturbed Hamiltonian has two degrees of freedom (our non-integrable pendulum is a special case of this example), the eigenvalue is given by
\begin{eqnarray}
  w^{(n_1, n_2)} (I_1, I_2)= n_1 \omega_1(I_1) + n_2 \omega_2(I_2) .
  \label{eq_eigenvalue2degree}
\end{eqnarray}
Thus, at the resonance, where the ratio of angular frequencies is $\omega_2(I_2) / \omega_1(I_1)= M / N$ with $M\not=0$ and $N\not=0$ being  coprime integers, the eigenvalue becomes zero when $n_2 / n_1=- N / M$.
Therefore, the eigenfunctions of the unperturbed Liouvillian are {\it infinitely degenerate} at specific values of the momenta in the generalized momenta space, since there is an 
 infinite number of combinations of $n_1$ and $n_2$ that satisfy this resonance condition. 

Let us make this infinite degeneracy for the eigenvalue problem of the unperturbed Liouvillian of our nonlinear pendulum more precise.
For this case the eigenvalue equation is written by
\begin{eqnarray}
  L_0\phi_0^{(n_1, n_2)} (I_1, \varphi_1, \varphi_2)=w^{(n_1, n_2)}(I_1) \phi_0^{(n_1, n_2)} (I_1, \varphi_1, \varphi_2), \quad
    \label{eq_evppho}
\end{eqnarray}
where
\begin{eqnarray}
  L_0&=&-i\left( \nu_1(I_1) \frac{\partial}{\partial \varphi_1} + \nu_2 \frac{\partial}{\partial \varphi_2} \right) ,
    \label{eq_jiouvillepho}
\end{eqnarray}
with
\begin{eqnarray}
  \nu_1(I_1)&=&\partial H_0/\partial I_1=2I_1,
    \label{eq_phoome1} \\
  \nu_2&=&\partial H_0/\partial I_2=1/\kappa .
    \label{eq_phoome2}
\end{eqnarray}
With arbitrary integers $n_1$ and $n_2$,
the eigenvalue and the eigenfunction 
are given by
\begin{eqnarray}
  w^{(n_1, n_2)}(I_1)&=& n_1 \omega_1(I_1) + n_2 \omega_2,
    \label{eq_ev0pho} \\
  \phi_0^{(n_1, n_2)} (I_1, \varphi_1, \varphi_2)&=&\frac{\sqrt{\Delta_1(I_1) \Delta_2}}{2\pi} e^{i(n_1 \Delta_1(I_1) \varphi_1 + n_2 \Delta_2 \varphi_2)},
    \label{eq_ef0pho}
\end{eqnarray}
with $\Delta_2= 1$.  
The angular frequencies $\omega_1(I_1)$ and $\omega_2$ are given by
\begin{eqnarray}
  \omega_1(I_1) &=& \Delta_1(I_1) \nu_1(I_1),
    \label{eq_phoome11} \\
  \omega_2 &=& \nu_2 ,
    \label{eq_phoome12}
\end{eqnarray}
with  $\Delta_1(I_1)$  defined at Eq.~(\ref{eq_kappa1}).
Then, we can write the eigenvalue equation for the unperturbed Liouvillian in the bra-ket notation as
\begin{eqnarray}
 L_0 | n_1,n_2 \rangle = w^{(n_1,n_2)}(I_1) | n_1,n_2\rangle. \label{eq.L0}
\end{eqnarray}

\begin{figure*}[t]
  \begin{center}
    \includegraphics[width=11.0cm,keepaspectratio]{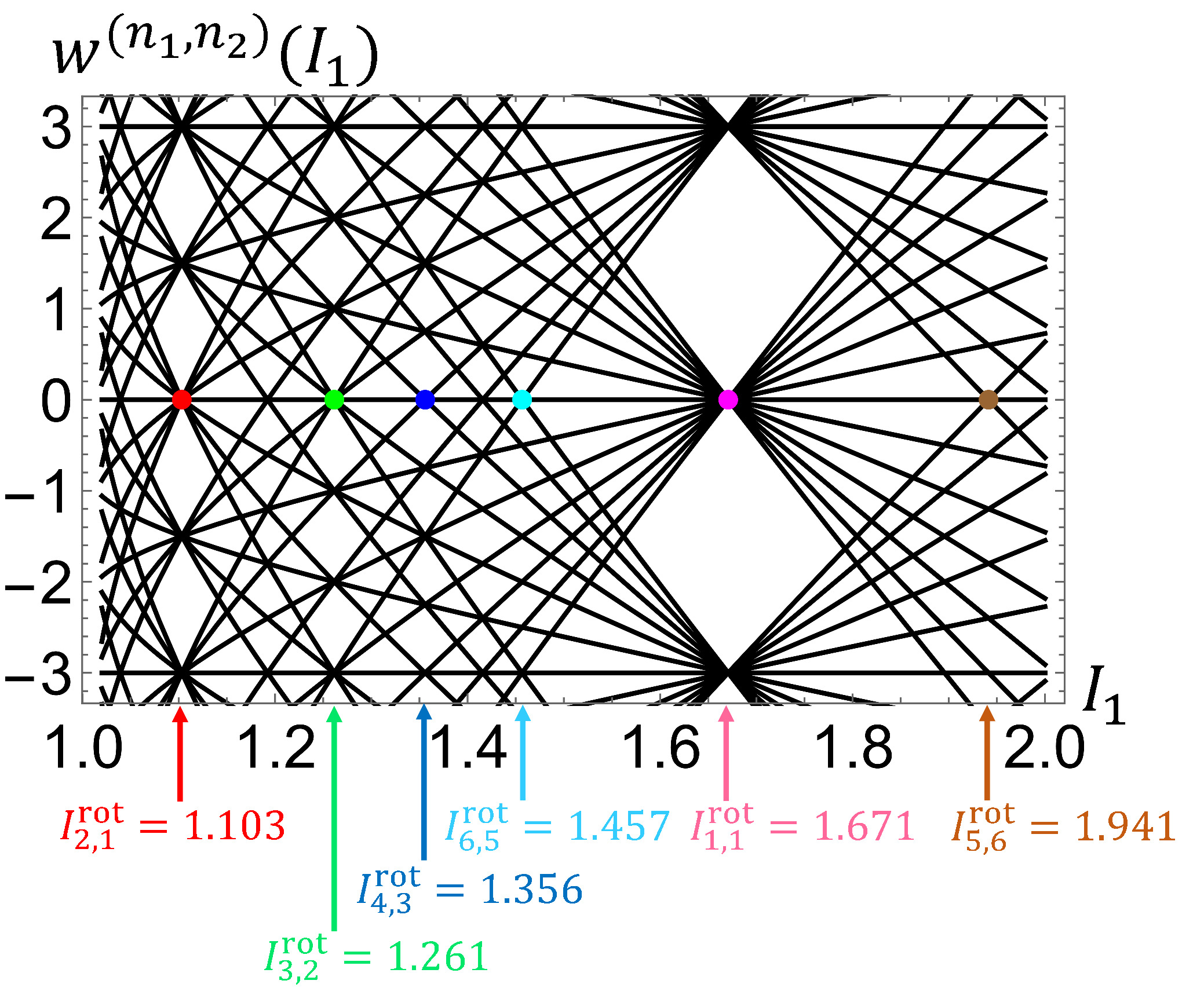}
    \caption{
Eigenvalues of the unperturbed Liouvillian of the non-integrable pendulum model shown in Eq.~(\ref{eq_ev0pho}) as a function of $I_1$ in the case of $\kappa=1/3$. A point at $I_1=I_{2,1}^{\rm rot}=1.103$ represents the $2/1$ resonance point, a point at $I_1=I_{3,2}^{\rm rot}=1.261$ is the $3/2$ resonance point, a points at $I_1=I_{4,3}^{\rm rot}=1.356$ is the $4/3$ resonance point, a point at $I_1=I_{6,5}^{\rm rot}=1.457$ is the $6/5$ resonance point, a points at $I_1=I_{1,1}^{\rm rot}=1.671$ is the $1/1$ primary resonance point, and a point at $I_1=I_{5,6}^{\rm rot}=1.941$ is the $5/6$ resonance point. We plot the eigenvalues for $|n_1|<6$ and $|n_2|<6$.
}
      \label{fig_eigenvalue0}
  \end{center}
\end{figure*}

Let us denote $I_1 =I_{M,N}$ that satisfies the $M / N$ resonance condition,
\begin{eqnarray}
\frac{\omega_2}{\omega_1(I_1)}=\frac{M }{N}, 
\label{resonanceCondition}
\end{eqnarray}
with $M\not=0$ and $N\not=0$ coprime integers.
At the resonance point $I_1 =I_{M,N}$, the eigenvalues are infinitely degenerate at zero for any integers $n_1$ and $n_2$ that satisfy $n_2 / n_1=- N / M$, i.e.,
\begin{eqnarray}
  w^{(\pm M, \mp N)} (I_{M,N})=
    w^{(\pm 2M, \mp 2N)} (I_{M,N}) &=&
    w^{(\pm 3M, \mp 3N)} (I_{M,N}) \nonumber \\
  &=& \cdots = 0 .
    \label{eq_degeneracypho}
\end{eqnarray}
In addition, the eigenvalue $w^{(0, 0)}$, which does not depend on $I_1$, is also zero.

We denote the value of $I_1$ as $I_{M,N}^{\rm rot}$  for the rotation of the pendulum and as $I_{M,N}^{\rm lib}$ for the libration of the pendulum at the $M/N$ resonance point. Fig. \ref{fig_eigenvalue0} shows the eigenvalue spectrum of the unperturbed Liouvillian in Eq.~(\ref{eq_ev0pho}) as a function of $I_1$.
Colored points in the figures represent degenerate zero eigenvalues at resonance points. 
The eigenvalues at each resonance point $I_1 = I_{M,N}$  are infinitely degenerate at $w^{(n_1, n_2)}(I_1)=...-3,0,3,6,...$ as shown in Fig. \ref{fig_eigenvalue0}.
The values of $I_1$ for
several resonance points are also shown on the figure of the Poincar\'e surface of section in Fig.~ \ref{fig_poincareivarphi}.

\section{Collision Operator}
\label{collisionoperator}

In this section, we will explain how to solve the eigenvalue problem of the total Liouvillian by addressing the nonlinear eigenvalue problem of the collision operator. This method is an extension of the method called the Brillouin-Wigner-Feshbach projection operator method \cite{Brillouin1932,Wigner1935,Feshbach1962} to solve the eigenvalue problem of the Hamiltonian operator in quantum mechanics.  We extend this method to solve the eigenvalue problem of the classical Liouvillian. Using this approach, we aim to estimate the magnitude of the frequency gap in the spectrum of the Liouvillian at the resonance point.

Let us introduce a projection operator with $\bm{n} = (n_1, n_2)$,
\begin{eqnarray}
  P^{(\bm{n})}
    &\equiv& | \bm{n} \rangle \langle \bm{n} |,
    \label{eq_projectionpn}
\end{eqnarray}
which projects onto an eigenstate $| \bm{n} \rangle$ of the unperturbed Liouvillian. This satisfies the relation,
\begin{eqnarray}
  L_0 P^{(\bm{n})} =w^{(\bm{n})}(\bm{I}) P^{(\bm{n})} . \label{eq_jp0}
\end{eqnarray}
We also introduce another projection operator 
\begin{eqnarray}
  Q^{(\bm{n})}
    &\equiv& \hat{I}_{\bm{\varphi}}-P^{(\bm{n})} \nonumber \\
    &=&\sum_{\quad \bm{n'}
   ( \ne \bm{n} )} | \bm{n'}\rangle \langle \bm{n'} | ,
    \label{eq_projectionqn}
\end{eqnarray}
which is complementary to $P^{(\bm{n})}$. These operators satisfy following relations,
\begin{eqnarray}
  L_0 P^{(\bm{n})}&=&P^{(\bm{n})} L_0 , \label{eq_jp} \\
  L_0 Q^{(\bm{n})}&=&Q^{(\bm{n})} L_0 \label{eq_jq}, \\
  \bigl(P^{(\bm{n})} \bigr)^2&=&P^{(\bm{n})} ,\label{eq_p2} \\
  \bigl(Q^{(\bm{n})} \bigr)^2&=&Q^{(\bm{n})}, \label{eq_q2} \\
  P^{(\bm{n})}+Q^{(\bm{n})}&=&\hat{I}_{\bm{\varphi}}, \label{eq_pqplus}\\
  P^{(\bm{n})}Q^{(\bm{n})}&=&Q^{(\bm{n})}P^{(\bm{n})}=0, \label{eq_pqtimes} \\
  \lambda P^{(\bm{n})} L_V P^{(\bm{n})}&=& 0,
   \label{eq_plvp}
\end{eqnarray}
where Eq.~(\ref{eq_plvp}) is obtained by Eq.~(\ref{eq_matrixelementprime1}) with $\lambda V_{\bm{0}} =0$.

Multiplying 
the eigenvalue equation
(\ref{eq_evp}) by $P^{(\bm{n})}$ and $Q^{(\bm{n})}$ from the left side, and using $P^{(\bm{n})}+Q^{(\bm{n})}=\hat{I}_{\bm{\varphi}}$, we obtain
\begin{eqnarray}
  P^{(\bm{n})} L_H (P^{(\bm{n})}+Q^{(\bm{n})}) | \Phi_{\alpha}^{(\bm{n})} \rangle=\Omega_{\alpha}^{(\bm{n})}(\bm{I}) P^{(\bm{n})} | \Phi_{\alpha}^{(\bm{n})} \rangle,
    \label{eq_evpbrap} \\
  Q^{(\bm{n})} L_H (P^{(\bm{n})}+Q^{(\bm{n})}) | \Phi_{\alpha}^{(\bm{n})} \rangle=\Omega_{\alpha}^{(\bm{n})}(\bm{I}) Q^{(\bm{n})} | \Phi_{\alpha}^{(\bm{n})} \rangle .
    \label{eq_evpbraq}
\end{eqnarray}
From Eq. (\ref{eq_evpbraq}) we obtain
\begin{eqnarray}
  Q^{(\bm{n})}| \Phi_{\alpha}^{(\bm{n})} \rangle
  =\frac{1}{\Omega_{\alpha}^{(\bm{n})}(\bm{I}) - Q^{(\bm{n})} L_H Q^{(\bm{n})}} Q^{(\bm{n})} L_H P^{(\bm{n})} | \Phi_{\alpha}^{(\bm{n})} \rangle . \nonumber \\
    \label{eq_evpbraq1}
\end{eqnarray}
Substituting Eq.~(\ref{eq_evpbraq1}) into Eq.~(\ref{eq_evpbrap}), we obtain an eigenvalue problem for 
the new operator $\psi^{(\bm{n})}(\Omega_{\alpha}^{(\bm{n})}(\bm{I}))$,
\begin{eqnarray}
  \psi^{(\bm{n})}(\Omega_{\alpha}^{(\bm{n})}(\bm{I})) P^{(\bm{n})} | \Phi_{\alpha}^{(\bm{n})} \rangle
  =\Omega_{\alpha}^{(\bm{n})}(\bm{I}) P^{(\bm{n})} | \Phi_{\alpha}^{(\bm{n})} \rangle,
    \label{eq_evpeff}
\end{eqnarray}
where
\begin{eqnarray}
  & &\hspace{-15pt} \psi^{(\bm{n})}(\Omega_{\alpha}^{(\bm{n})}(\bm{I})) \nonumber \\
  &\equiv&  P^{(\bm{n})} L_H P^{(\bm{n})} \nonumber \\
  && + P^{(\bm{n})} L_H Q^{(\bm{n})} \frac{1}{\Omega_{\alpha}^{(\bm{n})}(\bm{I}) - Q^{(\bm{n})} L_H Q^{(\bm{n})}} Q^{(\bm{n})} L_H P^{(\bm{n})}. \nonumber \\
    \label{eq_effl}
\end{eqnarray}
Here, $\psi^{(\bm{n})}$ is called {\it the collision operator}, which plays a central role in non-equilibrium statistical mechanics
\cite{Prigogine1961v2, Balescu1963,Petrosky1996, Petrosky1997}.

We note that the eigenvalue of the collision operator $\Omega_{\alpha}^{(\bm{n})}(\bm{I})$ in Eq. (\ref{eq_evpeff}) is the same as the eigenvalue of the Liouvillian in Eq. (\ref{eq_evp}). Furthermore, we note that the collision operator itself depends on its eigenvalue $\Omega_{\alpha}^{(\bm{n})}(\bm{I})$. In other words, Eq. (\ref{eq_evpeff}) is a nonlinear eigenvalue equation with respect to the eigenvalue. The eigenstate of the Liouvillian is written by the eigenstate and the eigenvalue of the collision operator as (see Eq. (\ref{eq_evpbraq1}))\cite{collision11}
\begin{eqnarray}
  | \Phi_{\alpha}^{(\bm{n})} \rangle 
    &=&P^{(\bm{n})}| \Phi_{\alpha}^{(\bm{n})} \rangle + Q^{(\bm{n})}| \Phi_{\alpha}^{(\bm{n})} \rangle \nonumber \\
    &=&\left( P^{(\bm{n})} + \frac{1}{\Omega_{\alpha}^{(\bm{n})}(\bm{I}) - Q^{(\bm{n})} L_H Q^{(\bm{n})}} Q^{(\bm{n})} L_H P^{(\bm{n})} \right) \nonumber \\
    & &\times P^{(\bm{n})} | \Phi_{\alpha}^{(\bm{n})} \rangle .
    \label{eq_evpbrapq0}
\end{eqnarray}

The collision operator is also called {\it the effective Liouvillian}, which corresponds to the effective Hamiltonian in quantum mechanics  \cite{Petrosky2010}. For the case of the Hamiltonian, the corresponding term to the second term in Eq. (\ref{eq_effl}) is called the self-energy part \cite{Mahan1990}. 
Hence, we can refer to the second term in Eq. (\ref{eq_effl}) as the self-frequency part of the effective Liouvillian.

\section{Frequency Gap in Liouvillian Spectrum}
\label{frequencygap}

Using the collision operator, let us now evaluate the frequency gap around the zero eigenvalue, which is the main part of this paper. Our objective is to determine the exponent $b$ such that the frequency gap is proportional to $\lambda^b$. Our idea is to evaluate the $\lambda$ dependence of the magnitude of this frequency gap by exploiting the fact that the eigenvalue problem of the collision operator above is nonlinear with respect to the eigenvalue, without explicitly solving the eigenvalue problem. 

In solving the problem, we use the method of perturbation expansion. For this purpose, we formally expand $1/(\Omega_{\alpha}^{(\bm{n})}(\bm{I}) - Q^{(\bm{n})} L_0 Q^{(\bm{n})} -\lambda Q^{(\bm{n})} L_V Q^{(\bm{n})})$ in the collision operator (\ref{eq_effl}) into a geometric series with a Neumann series in $\lambda Q^{(\bm{n})} L_V Q^{(\bm{n})}$ as,
\begin{widetext}
\begin{eqnarray}
 \psi^{(\bm{n})}(\Omega_{\alpha}^{(\bm{n})}(\bm{I})) &=& w^{(\bm{n})}(\bm{I}) P^{(\bm{n})} \nonumber \\
    &&+ \lambda^2 P^{(\bm{n})} L_V Q^{(\bm{n})} \left( \sum _{s=0}^\infty \left\{ R^{(\bm{n})}(\Omega_{\alpha}^{(\bm{n})}(\bm{I})) \right\}^s \right)  \frac{1}{\Omega_{\alpha}^{(\bm{n})}(\bm{I}) - Q^{(\bm{n})} L_0 Q^{(\bm{n})}}  Q^{(\bm{n})} L_V P^{(\bm{n})} , \qquad
      \label{eq_effl1}
\end{eqnarray}
\end{widetext}
where
\begin{widetext}
\begin{eqnarray}
  R^{(\bm{n})}(\Omega_{\alpha}^{(\bm{n})}(\bm{I}))
    &\equiv& \frac{\lambda}{\Omega_{\alpha}^{(\bm{n})}(\bm{I}) - Q^{(\bm{n})} L_0 Q^{(\bm{n})}} Q^{(\bm{n})} L_V Q^{(\bm{n})} \nonumber \\
    &=& \sum_{\bm{n'}} \sum_{ \bm{n''}}  \mid \hspace{-1pt} \bm{n'} \rangle \frac{\lambda}{\Omega_{\alpha}^{(\bm{n})}(\bm{I})-w^{(\bm{n'})}(\bm{I}) } \braket{ \bm{n'} \hspace{-1pt} \mid Q^{(\bm{n})}L_V Q^{(\bm{n})}\hspace{-1pt} \mid \hspace{-1pt} \bm{n''}}  \langle \bm{n''} \hspace{-1pt} \mid .  \hspace{20pt}\nonumber \\
    &=& \sum_{\quad \bm{n'} ( \ne \bm{n} )} \sum_{\quad \bm{n''} ( \ne \bm{n} )}
     \mid \hspace{-1pt} \bm{n'} \rangle \frac{\lambda}{\Omega_{\alpha}^{(\bm{n})}(\bm{I})-w^{(\bm{n'})}(\bm{I})} \braket{ \bm{n'} \hspace{-1pt} \mid L_V \hspace{-1pt} \mid \hspace{-1pt} \bm{n''}}  \langle \bm{n''} \hspace{-1pt} \mid .  \hspace{20pt}
    \label{eq_effl2}
\end{eqnarray}
\end{widetext}

In the following argument, we will restrict ourselves to the case where the coprime integers $M$ and $N$ for the $M/N$ resonance are not too large compared to 1. This is because, for large $N$, the resonance effect is proportional to a higher order of $\lambda$, as will be shown later in this section (see also the size of the islands in Figs.~\ref{fig_poincareivarphi} and \ref{fig_poincareivarphi40}), and for large $M$, the resonance effect is exponentially small due to the influence of the interaction in our system, as will also be shown later in this section.

To state the conclusion first, by examining the convergence condition of the collision operator, we can conclude that the lower bound on the $\lambda$ dependence of the frequency gap at the $M/N$ resonance is $\lambda^N$ in the case of our non-integrable pendulum.

The collision operator for the non-integrable system is written in the series by
\begin{eqnarray}
\psi^{(\bm{n})}(\Omega_{\alpha}^{(\bm{n})}(\bm{I}) )
= w^{(\bm{n})}(\bm{I}) \mid \hspace{-1pt} \bm{n} \rangle \langle \bm{n} |
    + \sum _{s=2}^\infty \lambda^{s} \psi_{s}^{(\bm{n})}(\Omega_{\alpha}^{(\bm{n})}(\bm{I})). \nonumber \\
      \label{eq_effl1appe}
\end{eqnarray}
For the case of our non-integrable pendulum, $w^{(\bm{n})}$ depends only on $I_1$, while $\Omega_{\alpha}^{(\bm{n})}(\bm{I})$ depends on both $I_1$ and $I_2$ (see Eq. \eqref{eq_ham23}).

In order to understand the structure of the collision operator, let us consider the forth order term  $\lambda^4 \psi_{4}^{(\bm{n})}$ as an example,
\begin{widetext}
\begin{eqnarray}
  \lambda^{4} \psi_{4}^{(\bm{n})} (\Omega_{\alpha}^{(\bm{n})}(\bm{I})) &=& \sum_{\bm{l}} \sum_{\bm{l'}} \sum_{\bm{l''}} \mid \hspace{-1pt} \bm{n} \rangle \nonumber \\
  &\times& \braket{ \bm{n} \hspace{-1pt} \mid P^{(\bm{n})} \lambda L_V Q^{(\bm{n})} \hspace{-1pt} \mid \hspace{-1pt} \bm{n} +\bm{l} } \frac{1}{\Omega_{\alpha}^{(\bm{n})}(\bm{I})-w^{( \bm{n} +\bm{l} )}(\bm{I})} \nonumber \\
  &\times& \braket{\bm{n} +\bm{l} \hspace{-1pt} \mid Q^{(\bm{n})} \lambda L_V Q^{(\bm{n})} \hspace{-1pt} \mid \hspace{-1pt} \bm{n} +\bm{l} + \bm{l'} } \frac{1}{\Omega_{\alpha}^{(\bm{n})}(\bm{I})-w^{( \bm{n} +\bm{l} + \bm{l'} )}(\bm{I})}  \nonumber \\ 
  &\times& \braket{ \bm{n} +\bm{l} + \bm{l'} \hspace{-1pt} \mid Q^{(\bm{n})}  \lambda L_V Q^{(\bm{n})} \hspace{-1pt} \mid \hspace{-1pt}  \bm{n} +\bm{l} + \bm{l'} + \bm{l''} } \frac{1}{\Omega_{\alpha}^{(\bm{n})}(\bm{I})-w^{( \bm{n} +\bm{l} + \bm{l'} + \bm{l''})}(\bm{I})}  \nonumber \\     
  &\times& \braket{\bm{n} +\bm{l} + \bm{l'}  + \bm{l''} \hspace{-1pt} \mid Q^{(\bm{n})}  \lambda L_V  P^{(\bm{n})} \hspace{-1pt} \mid \hspace{-1pt} \bm{n} }\langle \bm{n} |.
      \label{colli4}
\end{eqnarray}
\end{widetext}
Here, we denote the dummy variables appearing in the intermediate states by $\bm{l}$, $\bm{l'}$ and $\bm{l''}$, in order to distinguish them from the index $\bm{n}$ in the initial and final states.

We consider an $M/N$ resonance at $I_1 = I_{M,N}$ that satisfies the resonance condition ${\omega_2}/{\omega_1(I_1)}={M}/{N}$. In this case, as shown in Eq.~(\ref{eq_degeneracypho}), the eigenfunctions of the unperturbed Liouvillian are infinitely degenerate at the zero eigenvalue as $w^{(n_1,n_2)}(I_{M,N})=0$ for any non-zero integers $n_1$ and $n_2$ satisfying $n_2/n_1 = -N/M$ and also for  $n_1 =n_2 =0$. We are interested in how the perturbation affects the invariant of motion of the unperturbed system, which corresponds to the zero eigenvalue of the unperturbed Liouvillian. Thus, we focus our attention
on the eigenvalues of the perturbed Liouvillian associated with the zero eigenvalue of the unperturbed Liouvillian. 

Since we consider the case $\lambda \ll 1$, we are looking for the perturbed eigenvalue
\begin{eqnarray}
\Omega_\alpha^{(\bm{n})}(I_{M,N}, I_2) \propto \lambda^b
\quad {\rm with }\,\,\, \bm{n} = (m M,-m N), 
\label{Ome_b}
\end{eqnarray}
where  $M\not=0$ and $N\not=0$ are coprime integers, $b >0$, and $m$ is any integer including $m=0$.

Let us first consider  the contribution from the  interaction part $\lambda L_V$ in Eq. \eqref{colli4}. Note that in the case of the non-integrable pendulum considered here, the potential $V_{\bm{l}}(\bm{I})$ decreases exponentially when $|\bm{l}|$ becomes sufficiently larger than 1 because this Fourier coefficient is proportional to the inverse of the hyperbolic sine function as shown in Eq.~(\ref{eq_vfourierexpansion}). In fact, this exponential decrease holds not only for this model, but for a very wide class of potentials \cite{Barrar1970}. Therefore, the contribution that comes from these interactions does not lead to divergence of the collision operator when we take the summation over the dummy variables $\bm{l}$, $\bm{l'}$, ... appearing in the intermediate states. Hence, if the collision operator diverges, it comes from the propagators with a small denominator, such as $\Omega_{\alpha}^{(\bm{n})}(I_{M,N}, I_2)-w^{( \bm{n} +\bm{l})}(I_{M,N})$, and so on in Eq. (\ref{colli4}).

Therefore, we next focus our attention on the denominators in Eq. (\ref{colli4}). In order to find out the magnitude of the contribution from the denominators,
let us express the eigenvalue of the perturbed Liouvillian as
\begin{eqnarray}
 \Omega_\alpha^{(\bm{n})}(I_{M,N}, I_2) =\frac{\omega_2 C(I_{M,N}, I_2)}{M}, \label{eq.C}
\end{eqnarray}
where $C(I_{M,N}, I_2)$ is a dimensionless quantity that is proportional to $\lambda^b$ with $b>0$.

The unperturbed eigenvalue in the intermediate states in Eq. (\ref{colli4}) is, for example, given by
$w^{(\bm{n}+\bm{l})}(\bm{I}) 
= (n_1 + l_1)\omega_1(I_1) + (n_2 + l_2)\omega_2$.
Hence, for $I_1 =I_{M,N}$ resonance point with ${\omega_2}/{\omega_1(I_1)}={M}/{N}$, we have
\begin{eqnarray}
  & &\Omega_{\alpha}^{(\bm{n})}({I}_{M,N}, I_2)- w^{(\bm{n}+\bm{l})}(I_{M,N}) \nonumber \\
  & &\hspace{30pt}= \frac{\omega_2}{M} \Bigl[   C(I_{M,N}, I_2) - (n_1 + l_1)N  - (n_2 + l_2)M \Bigr]. \quad
\label{wMNdenominator}
\end{eqnarray}
We note 
\begin{eqnarray}
(n_1 + l_1)N + (n_2 + l_2)M =  l_1N + l_2M,
\label{wMNdenominator2}
\end{eqnarray}
because of $n_2/n_1 = -N/M$ or $n_1 =n_2 =0$. The important point here is that the values $l_1N + l_2M$, $(l_1 + l_1')N + (l_2 + l_2')M$ and $(l_1 + l_1' + l_1'')N + (l_2 + l_2' + l_2'')M$ in the intermediate states can only be an integer.  
Moreover, we should note that even with the restriction $Q^{(\bm{n})}$ on $\bm{n} = (mM,-mN)$ in the intermediate states, these integers may be 0 for the dummy variables $\bm{l}$, $\bm{l'}$ and $\bm{l''}$ that satisfy  $l_2 /l_1= -N/M$, or $(l_2 + l_2') / (l_1 + l_1') = -N/M$, or $ (l_2 + l_2'+ l_2'') / (l_1 + l_1'+ l_1'')= -N/M$.

Based on the above observations, 
let us first consider  
the effect of the interaction at the $N$-th order resonance point. Then, from Eq. \eqref{colli4} we have
\begin{widetext}
\begin{eqnarray}
  &&\hspace{-20pt} \lambda^{4} \psi_{4}^{(\bm{n})} \bigl(\Omega_{\alpha}^{(\bm{n})} \bigr)\nonumber \\
  &=& \sum_{\bm{l}} \sum_{\bm{l'}} \sum_{\bm{l''}} \mid \hspace{-1pt} \bm{n} \rangle \braket{ \bm{n} \hspace{-1pt} \mid 
        P^{(\bm{n})} \lambda L_V Q^{(\bm{n})} \hspace{-1pt} \mid \hspace{-1pt} \bm{n} +\bm{l} } \frac{1}{\frac{\omega_2}{M} \bigl[ C -  l_1 N -  l_2M \bigr] } \nonumber \\
  & &\hspace{20pt} \times \braket{\bm{n} +\bm{l} \hspace{-1pt} \mid Q^{(\bm{n})} \lambda L_V Q^{(\bm{n})} \hspace{-1pt} \mid \hspace{-1pt} \bm{n} +\bm{l} + \bm{l'} } \frac{1}{\frac{\omega_2}{M} \bigl[  C - (l_1 + l'_1 ) N  - (l_2+ l'_2 ) M\bigr]  } \nonumber \\
  & &\hspace{20pt} \times \braket{\bm{n} +\bm{l} + \bm{l'} \hspace{-1pt} \mid Q^{(\bm{n})} \lambda L_V Q^{(\bm{n})} \hspace{-1pt} \mid \hspace{-1pt} \bm{n} +\bm{l} + \bm{l'}+ \bm{l''} } \frac{1}{\frac{\omega_2}{M} \bigl[  C - (l_1 + l'_1  + l''_1) N  - (l_2+ l'_2  + l''_2) M\bigr]  } \nonumber \\
  & &\hspace{20pt} \times \braket{\bm{n} +\bm{l} + \bm{l'} + \bm{l''}\hspace{-1pt} \mid Q^{(\bm{n})} \lambda L_V P^{(\bm{n})} \hspace{-1pt} \mid \hspace{-1pt} \bm{n} }\langle \bm{n}|,
      \label{colli42}
\end{eqnarray}
\end{widetext}
where we have dropped the arguments $(I_{M,N}, I_2)$ in $\Omega_{\alpha}^{(\bm{n})}$ and $C$ to avoid too heavy notations.

Due to the specific form of the interaction in Eq. (\ref{eq_ham13}), the Fourier coefficient of the interaction, $V_{l_1, l_2}(I_1, I_2)$, with respect to $\varphi_2$ has only the Fourier index with $l_2= \pm1$ as shown in Eq.~(\ref{eq_vfourierexpansion00}). Hence, due to the matrix element of the perturbed Liouvillian in Eq.~(\ref{eq_matrixelementprime1}), the possible values in the summation of Eq.~(\ref{colli42}) are $l_2 = \pm 1$,  $l_2+ l_2' = \pm 2$ or $0$,  and $l_2+ l_2'+l_2'' = \pm 3$ or $\pm 1$. However, because of the factor $Q^{(\bm{n})}$ in the intermediate state, the case $l_2+ l_2'= 0$ are excluded when $l_1+ l_1' = 0$.
With this observation, let focus our attention on the primary resonance 
 with $N=1$.
For this case,
one may reach the intermediate state with the zero eigenvalue of the unperturbed Liouvillian through a single transition with $\lambda L_V$. Indeed, $l_1 = \mp M$,  $l_1+ l_1' = \mp 2 M$, $l_1+ l_1'+l_1'' = \mp 3 M$ or $\mp  M$ are these cases where $l_1 N +l_2 M =0$, $(l_1 +l_1')N +(l_2 +l_2')M =0$, and $(l_1 +l_1' + l_1'')N +(l_2 +l_2' + l_2'')M =0$ in the denominators in Eq. \eqref{colli42}. Denoting these special values of the dummy variables for the resonance condition as $\bm{l}_r$, $\bm{l'}_r$, and so on, we note that these special values satisfy
\begin{eqnarray}
\frac{l_{r,2}+ l_{r,2}'+ \cdots + l_{r,2}^{[k]} } { l_{r,1}+ l_{r,1}'+ \cdots + l_{r,1}^{[k]} }= -\frac{N} {M} 
\label{lrk}
\end{eqnarray}
with $N=1$ (see Appendix \ref{section_C}).

Therefore, for the case of  $C \ll 1$, as explained just below Eq. \eqref{wMNdenominator2}, the largest contributing term in this $\lambda^4$ order contribution comes from these special values when summing the dummy variables, and it has the following form:
\begin{eqnarray}
  & &\hspace{-20pt} \Bigl[ D_{M/N, 4} \Bigr]_{N=1} \nonumber \\
  &=&
        \mid \hspace{-1pt} \bm{n} \rangle \braket{ \bm{n} \hspace{-1pt} \mid 
        P^{(\bm{n})} L_V Q^{(\bm{n})}
     \hspace{-1pt} \mid \hspace{-1pt} \bm{n} +\bm{l}_r  }
\frac{\lambda }{\frac{\omega_2}{M} 
   C
  } \nonumber \\
   &&\times
    \braket{\bm{n} +\bm{l}_r  \hspace{-1pt} \mid 
   Q^{(\bm{n})}  L_V Q^{(\bm{n})}
    \hspace{-1pt} \mid \hspace{-1pt} \bm{n} +\bm{l}_r  + \bm{l'}_r  }
    \frac{\lambda }{\frac{\omega_2}{M}
      C
      } \nonumber \\
   &&\times
    \braket{\bm{n} +\bm{l}_r + \bm{l'}_r  \hspace{-1pt} \mid 
   Q^{(\bm{n})} L_V Q^{(\bm{n})}
    \hspace{-1pt} \mid \hspace{-1pt} \bm{n} +\bm{l}_r  + \bm{l'}_r + \bm{l''}_r   }
    \frac{\lambda }{\frac{\omega_2}{M} 
      C
     } \nonumber \\
    &&\times
    \braket{\bm{n} +\bm{l}_r  + \bm{l'}_r  + \bm{l''}_r \hspace{-1pt} \mid 
    Q^{(\bm{n})} \lambda L_V P^{(\bm{n})}
    \hspace{-1pt} \mid \hspace{-1pt} \bm{n} }\langle \bm{n}| .
      \label{D4N1}
\end{eqnarray}

From this structure it is easy to understand that the largest term $ D_{M/1, j}$ of 
$\lambda^{j} \psi_{j}^{(\bm{n})} (\Omega_{\alpha}^{(\bm{n})})$ for $\lambda \ll 1$  with $N=1$ for any integer $j>1$ is
\begin{eqnarray}
D_{M/1, j} \propto \Bigl[ \frac{\lambda M}{C(I_{M,1}, I_2)} \Bigr]^{j-1} \lambda.
       \label{DM1j}
\end{eqnarray}
(Recall the lowest order contribution from the interaction starts form $\lambda^2$  in the series expansion of the collision operator in $\lambda$.) Hence, for $C(I_{M,1}, I_2) \propto \lambda^b$, $b=1$ is the border of convergence of the collision operator for the primary resonance when we take the summation over $j$ to the infinity. That is, for $b<1$ the collision operator converges, and for $b>1$ the collision operator diverges at the primary resonance. As a result, we can conclude that the lower bound of the frequency gap at the primary resonance with $N=1$ is proportional to $\lambda$.

Before going to the general case of $N$, we next consider the effect of the interaction to the secondary resonance with $N = 2$. Again we consider the expression \eqref{colli42}. Let us read this expression from left to right. Because  $l_{r, 2} = \pm 1$, we can see from Eq.~(\ref{lrk}) that for $l_{r, 1} = \pm M/N$ with $N=2$ the denominator of the first propagator from the left reduces to the small value $(\omega_2/M)C$. However, it is impossible because $l_{r, 1}$ cannot be an integer since $M$ and $N=2$ are coprime integers. Namely, there are no combinations of integers $l_{r, 1}$ and $l_{r, 2}$ that satisfy the resonance condition (\ref{lrk}).

 On the other hand, since $l_{r, 2} + l_{r, 2}'= \pm 2$, the resonance condition $(l_{r, 2} + l_{r, 2}')/(l_{r, 1} + l_{r, 1}') = -N/M$ for $N=2$ leads to $l_{r, 1}+ l_{r, 1}' = \pm M$ which are integers. Hence, the denominator of the second propagator from the left can reduce to a small value $(\omega_2/M)C$. This shows that there must be at least two transitions of $\lambda L_V$, from the leftmost state $| \bm{n} \rangle$ to the intermediate states associated with the $M/N$ resonance condition for $N = 2$. As a result,  the largest contributing term in this $\lambda^4$ order contribution of Eq.~\eqref{colli4} for $N=2$ has the following form:
\begin{eqnarray}
  & &\hspace{-20pt} \Bigl[ D_{M/N, 4} \Big]_{N=2} \nonumber \\
  &=&
   \mid \hspace{-1pt} \bm{n} \rangle \braket{ \bm{n} \hspace{-1pt} \mid 
 P^{(\bm{n})} \mathcal{L}_{V^2}^{(\bm{n})}\bigl(\Omega^{(\bm{n})}_\alpha \bigr) Q^{(\bm{n})}  
   \hspace{-1pt} \mid \hspace{-1pt} \bm{n} +\bm{l}_r }
\frac{ \lambda^2}{\frac{\omega_2}{M} C(I_{M,2}, I_2)  } \nonumber \\
   & &\times 
    \braket{\bm{n} +\bm{l}_r  \hspace{-1pt} \mid  
Q^{(\bm{n})}\lambda^2\mathcal{L}_{V^2}^{(\bm{n})}\bigl(\Omega^{(\bm{n})}_\alpha \bigr) P^{(\bm{n})}
    \hspace{-1pt} \mid \hspace{-1pt} \bm{n} }\langle \bm{n}|,    
      \label{DMN4}
\end{eqnarray}
where a  {\it renormalized interaction} is defined by
\begin{eqnarray}
 \hspace{-30pt} 
 \lambda^2\mathcal{L}_{V^2}^{(\bm{n})}(\Omega)
  \equiv 
    \lambda L_V Q^{(\bm{n})}
   \frac{1}{\Omega
   - L_0  }
     Q^{(\bm{n})}\lambda L_V .    
      \label{eq_LV2}
\end{eqnarray}

Similarly, we have the following form for the largest term $ D_{M/2, 6}$ of 
$\lambda^{6} \psi_{6}^{(\bm{n})} (\Omega_{\alpha}^{(\bm{n})})$ with $N=2$ in the $\lambda^6$ order contribution,
\begin{eqnarray}
  & &\hspace{-15pt} \Bigl[ D_{M/N, 6} \Big]_{N=2} \nonumber \\
  &=&
   \mid \hspace{-1pt} \bm{n} \rangle \braket{ \bm{n} \hspace{-1pt} \mid 
 P^{(\bm{n})} \mathcal{L}_{V^2}^{(\bm{n})}\bigl(\Omega^{(\bm{n})}_\alpha \bigr) Q^{(\bm{n})}  
   \hspace{-1pt} \mid \hspace{-1pt} \bm{n} +\bm{l}_r }
\frac{ \lambda^2}{\frac{\omega_2}{M} C(I_{M,2}, I_2)  } \nonumber \\
     & &\times
 \braket{\bm{n} +\bm{l}_r \hspace{-1pt} \mid 
Q^{(\bm{n})}\mathcal{L}_{V^2}^{(\bm{n})}\bigl(\Omega^{(\bm{n})}_\alpha \bigr) Q^{(\bm{n})}
    \hspace{-1pt} \mid \hspace{-1pt} \bm{n} +\bm{l}_r + \bm{l'}_r }
 \frac{ \lambda^2}{\frac{\omega_2}{M} C(I_{M,2}, I_2)  } \nonumber \\
    & &\times
    \braket{\bm{n} +\bm{l}_r  + \bm{l'}_r  \hspace{-1pt} \mid  
Q^{(\bm{n})}\lambda^2\mathcal{L}_{V^2}^{(\bm{n})}\bigl(\Omega^{(\bm{n})}_\alpha \bigr) P^{(\bm{n})}
    \hspace{-1pt} \mid \hspace{-1pt} \bm{n} }\langle \bm{n}|.
      \label{DMN6}
\end{eqnarray}

From this structure it is again easy to understand that the largest term $ D_{M/2, j}$ of 
$\lambda^{2j} \psi_{2j}^{(\bm{n})} (\Omega_{\alpha}^{(\bm{n})}) $ for $\lambda \ll 1$  with $N=2$ for any integer $j>1$ is
\begin{eqnarray}
D_{M/2, 2j} \propto \Bigl[ \frac{\lambda^2 M}{C(I_{M,2}, I_2)} \Bigr]^{j-1} \lambda^2.
       \label{DM2j}
\end{eqnarray}
Hence, for $C(I_{M,2}, I_2) \propto \lambda^b$, $b=2$ is the border of convergence of the collision operator at the secondary resonance.  As a result, we can conclude that the lower bound of the frequency gap at the secondary resonance with $N=2$ is proportional to $\lambda^2$.

Repeating a similar argument for any value of $N$, one can show that we need at least $N$ times transitions  of $\lambda L_V$ from the state $| \bm{n} \rangle$ to the intermediate state associated with the $M/N$ resonance condition with any value of $N$.  However, for $N\ge 3$ we have to use the following more general expression of the renormalized interaction than Eq.~\eqref{eq_LV2},
\begin{eqnarray}
  & &\hspace{-20pt} 
  \lambda^N\mathcal{L}_{V^N}^{(\bm{n})}(\Omega) \nonumber \\
  &\equiv& 
    \lambda L_V Q^{(\bm{n})}
    \Bigl[ 
   \frac{1}{\Omega
   - L_0  }
     Q^{(\bm{n})}\lambda L_V Q^{(\bm{n})}
     \Bigr]^{N-2} 
   \frac{1}{\Omega
   - L_0  }
     Q^{(\bm{n})}\lambda L_V . \nonumber \\
      \label{eq_LVN}
\end{eqnarray}
Then, one can show that the largest term $ D_{M/N, Nj}$ of 
$\lambda^{Nj} \psi_{Nj}^{(\bm{n})} (\Omega_{\alpha}^{(\bm{n})})$ for $\lambda \ll 1$  with any $N$ and for any integer $j>1$ is
\begin{eqnarray}
D_{M/N, Nj} \propto \Bigl[ \frac{\lambda^N M}{C(I_{M,N}, I_2)} \Bigr]^{j-1} \lambda^N.
       \label{DMNj}
\end{eqnarray}
Hence, for $C(I_{M,N}, I_2) \propto \lambda^b$, $b=N$ is the border of convergence of the collision operator at the $M/N$ resonance with any value of $N$. As a result, we can conclude that the lower bound of the frequency gap at the $N$-th order resonance is proportional to $\lambda^N$.
In Appendix \ref{section_C}, we present a proof of Eq. \eqref{DMNj}.

In addition, we can show that there is an exception among the eigenfunctions with a zero eigenvalue of the Liouvillian, where the denominators in Eq.~(\ref{colli4}) exactly cancel with the numerators when the collision operator acts on a special eigenstate. This is the case where the special eigenstate is a function of only the Hamiltonian. We will discuss this exceptional case in Section \ref{ResonanceAndHamiltonian}.

\section{Comparison with Numerical Simulation}
\label{Numerical}

In this section, we compare theoretical estimates of the frequency gap at the resonance points with results obtained by numerical calculation of the equations of motion for the trajectories in phase space.

Using the eigenvalue $\Omega_{\alpha}^{(\bm{n})}(\bm{I})$ and the eigenstate $\mid \hspace{-3pt} \Phi_{\alpha}^{(\bm{n})} \rangle$ of the perturbed Liouvillian in Eq. (\ref{eq_evp1}), we obtain the solution of the time evolution of a state function in the phase space as
\begin{eqnarray}
  | \rho(\bm{I}, t) \rangle &=& e^{-iL_Ht} | \rho(\bm{I}, 0) \rangle \nonumber \\
    &=&\sum_{\bm{n}} \sum_{\alpha} e^{-iL_Ht} | \Phi_{\alpha}^{(\bm{n})} \rangle \braket{\Phi_{\alpha}^{(\bm{n})} |  \rho(\bm{I}, 0) } \nonumber \\
    &=&\sum_{\bm{n}} \sum_{\alpha} e^{-i\Omega_{\alpha}^{(\bm{n})}(\bm{I})t} | \Phi_{\alpha}^{(\bm{n})} \rangle  W_{\alpha}^{(\bm{n})}(\bm{I}), 
      \label{eq_rhoformalbraket}
\end{eqnarray}
where $W_{\alpha}^{(\bm{n})}(\bm{I})\equiv \braket{\Phi_{\alpha}^{(\bm{n})} | \rho(\bm{I}, 0) }$ is the weight function that the initial condition of the state function is found in an eigenmode of the total Liouvillian. Then we have
\begin{eqnarray}
  \rho(\bm{I}, \bm{\varphi}, t)
    &\equiv& \braket{\bm{\varphi} | \rho(\bm{I}, t) } \nonumber \\
    &=& \sum_{\bm{n}} \sum_{\alpha} \Phi_{\alpha}^{(\bm{n})} (\bm{I}, \bm{\varphi}) e^{-i\Omega_{\alpha}^{(\bm{n})}(\bm{I})t} W_{\alpha}^{(\bm{n})}(\bm{I}) .
    \label{eq_rhoformal}
\end{eqnarray}

We note that when the $\delta$ function in the phase space is expanded by the eigenmode of the evolution operator, the weight function associated with each mode is more or less the same order of magnitude. We also note that a time evolution of the  $\delta$ function in the phase space represents the evolution of a trajectory. Hence, we can expect that if the Fourier transformation of the time evolution of the trajectory is analyzed, we can find the possible value of the angular frequency around the resonance point. Therefore, we can expect that this analysis will confirm the validity of our theoretical estimation of the frequency gap by examining how the lower bound of the possible frequencies for a given trajectory behaves as a function of $\lambda$ for $\lambda \ll 1$. Hence, we have examined the distribution of the frequency spectrum of dynamical variables for various trajectories in order to identify the order of magnitude of the frequency gap. As our system is non-integrable, we will investigate the frequency spectra by solving the Hamilton's equations of motion numerically.

\begin{figure*}[t]
  \begin{center}
    \begin{tabular}{c}
        \begin{minipage}{0.50\hsize}
        \begin{center}
          \includegraphics[height=4cm,keepaspectratio]{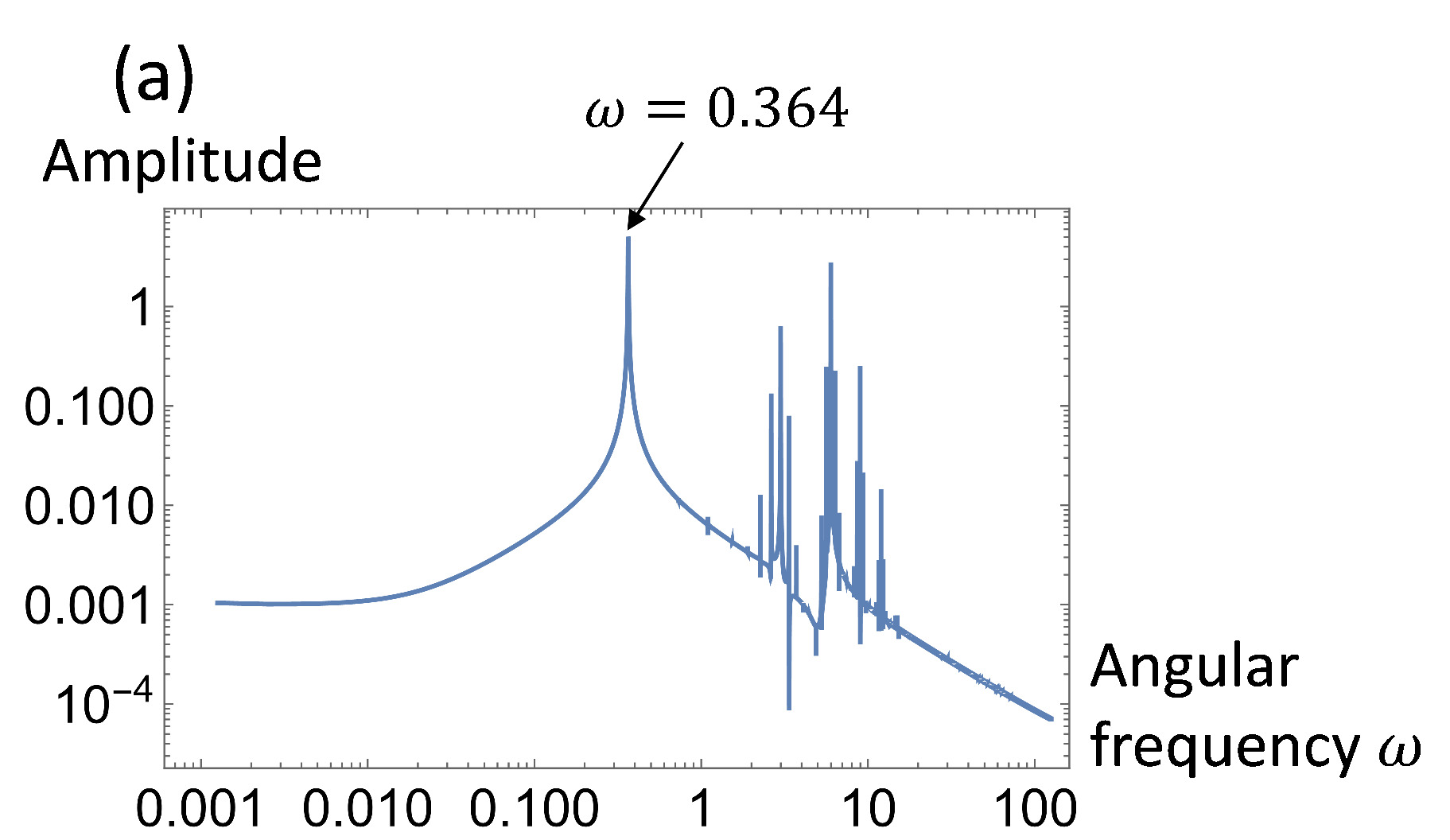}
        \end{center}
      \end{minipage}
        \begin{minipage}{0.50\hsize}
        \begin{center}
          \includegraphics[height=4cm,keepaspectratio]{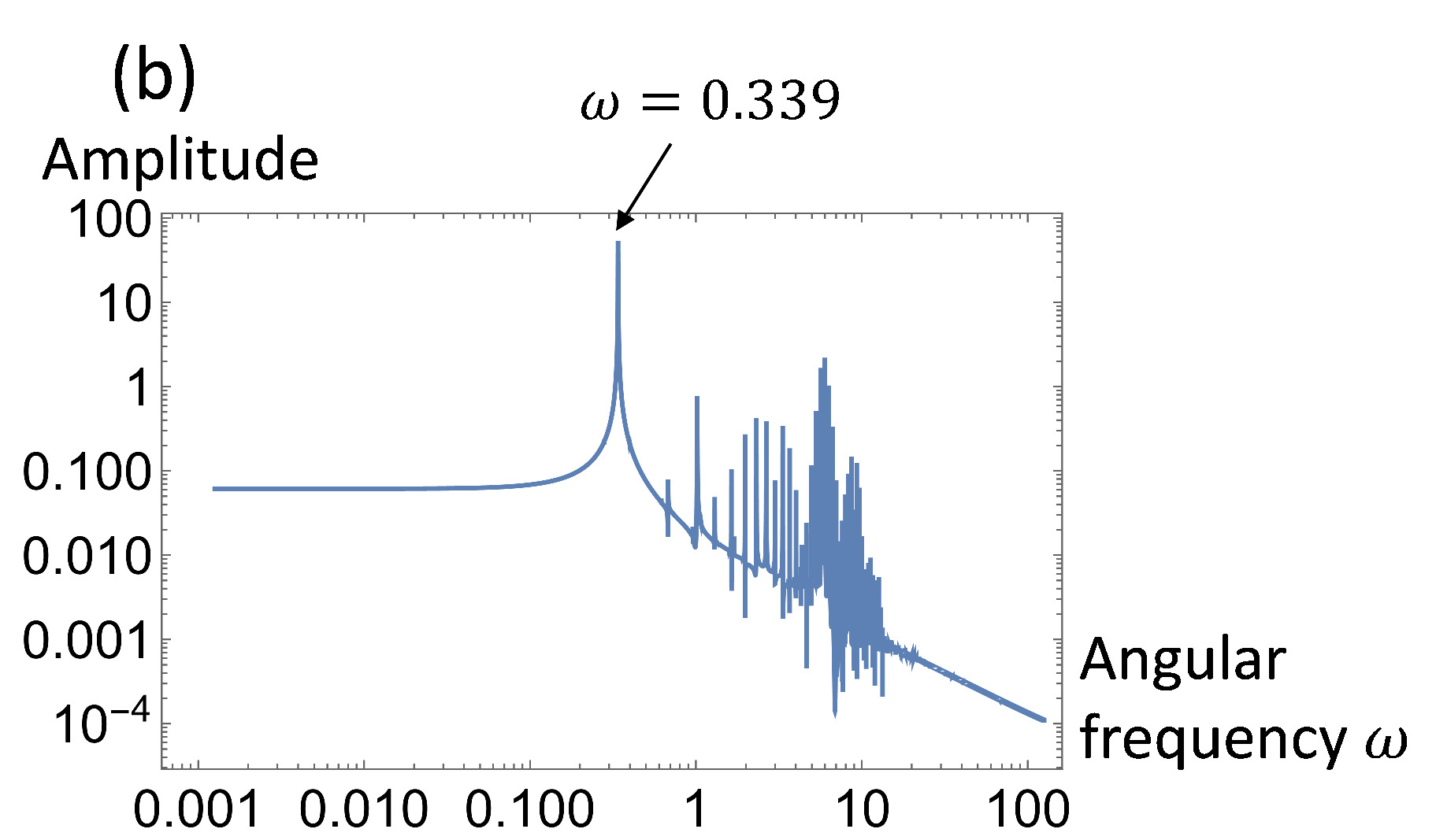}
        \end{center}
      \end{minipage}\\ \\
        \begin{minipage}{0.50\hsize}
        \begin{center}
          \includegraphics[height=4cm,keepaspectratio]{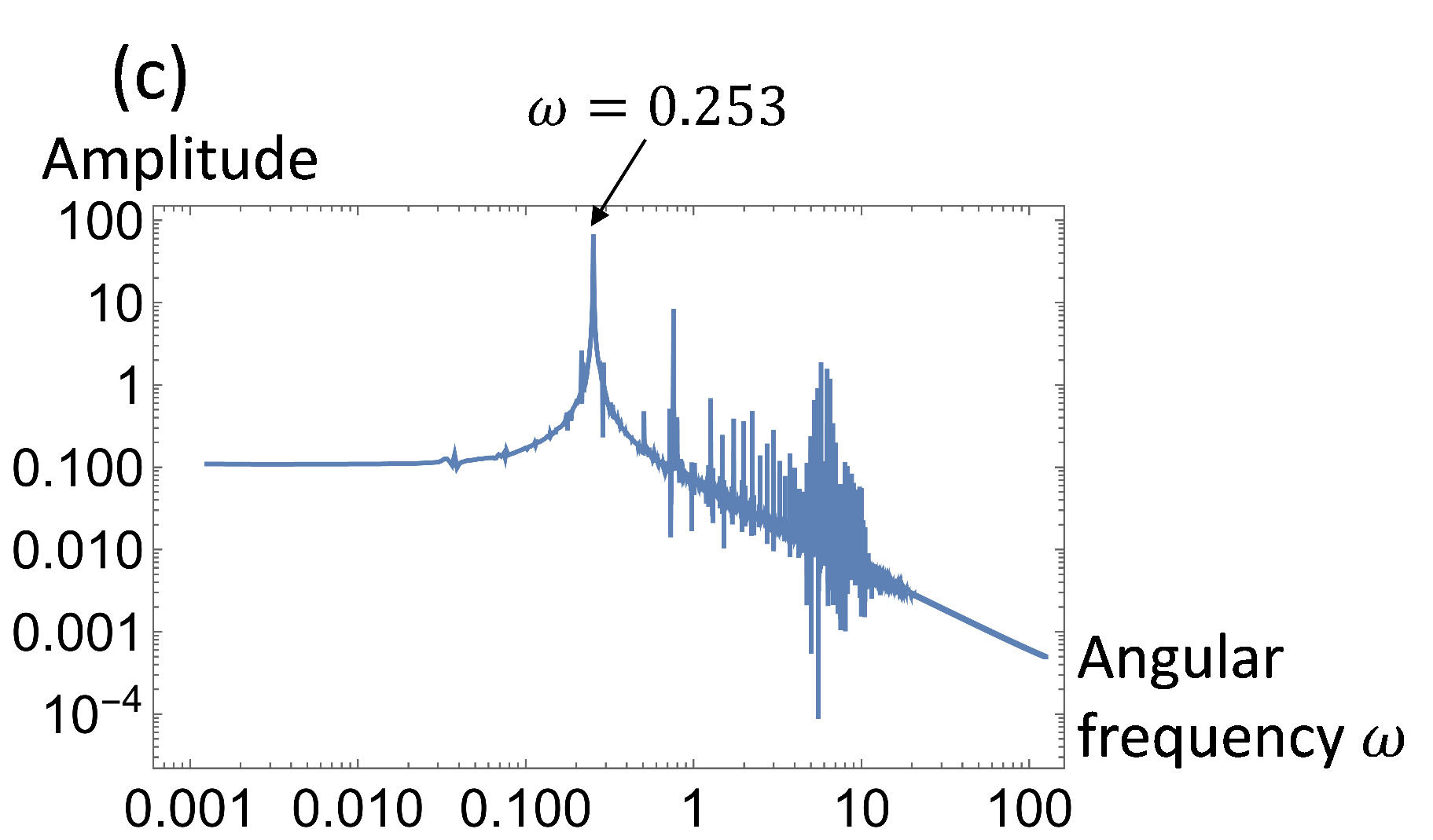}
        \end{center}
      \end{minipage}
        \begin{minipage}{0.50\hsize}
        \begin{center}
          \includegraphics[height=4cm,keepaspectratio]{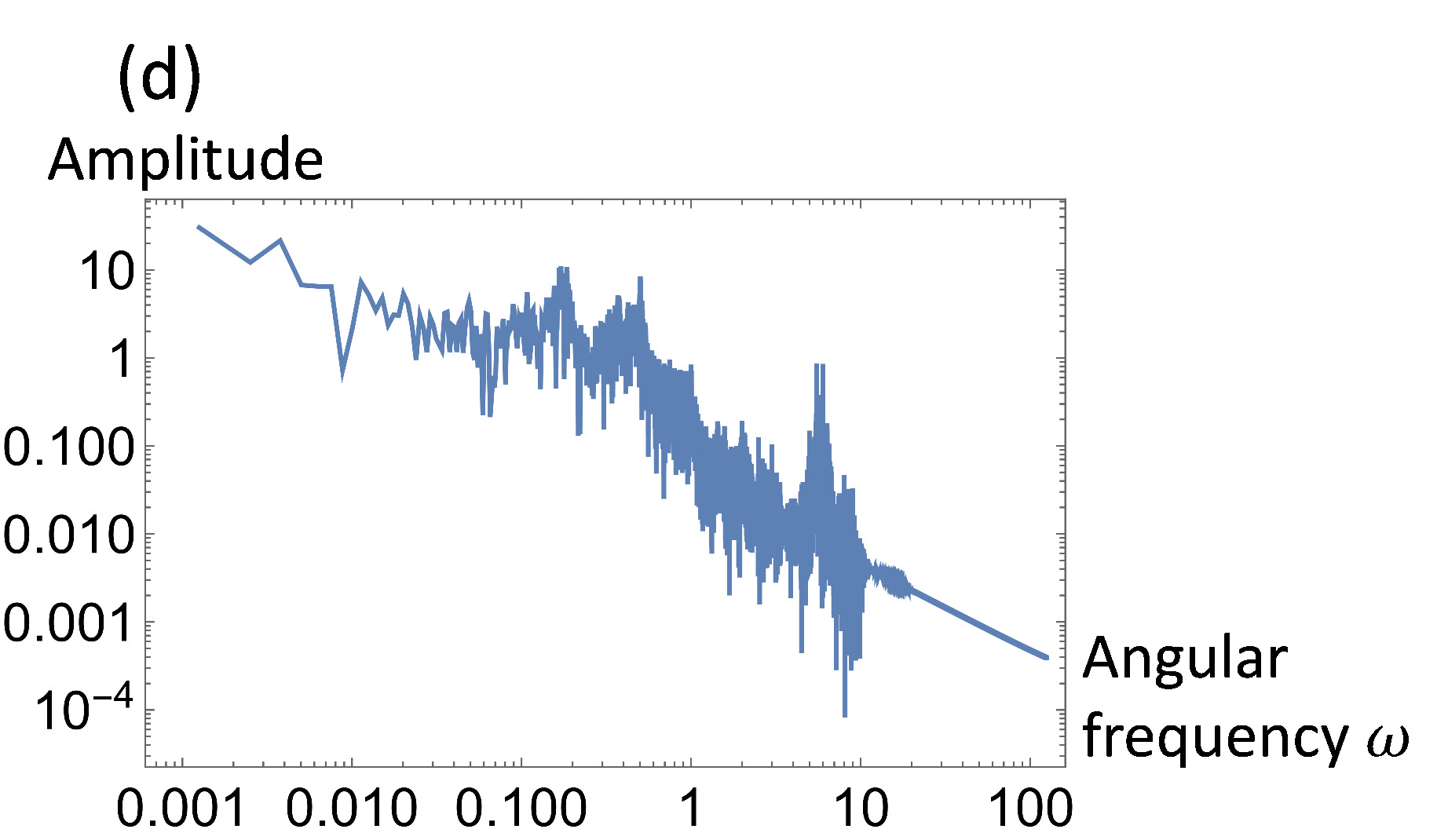}
        \end{center}
      \end{minipage}
    \end{tabular}
\caption{
Fourier spectra of $I_1(\tau)$ for trajectories around the $1/1$ primary resonance point with respect to time $\tau$ for the system with the dimensionless Hamiltonian $H=18.222$ in the case of $\lambda=1/10$ and $\kappa = 1/3$. The Fourier transformation is performed over the time interval $\tau=0$ to $5000$. Figures (a), (b), (c), and (d) depict the Fourier spectra of trajectories corresponding to (i), (ii), (iii), and (iv) on the Poincar\'e surface of section in Fig. \ref{fig_poincareivarphi}, respectively.
Note that trajectories corresponding to (i), (ii), and (iii) exhibit almost regular motion, whereas trajectory corresponding to (iv) exhibits a chaotic motion, as shown in Fig. \ref{fig_poincareivarphi}.
}
\label{fig_fourieranalysis}
  \end{center}
\end{figure*}

In Fig. \ref{fig_fourieranalysis}, we show some results of the distributions of the frequency spectra obtained from the Fourier transformation of the time evolution of the generalized momentum $I_1(\tau)$ with the continuous time $\tau$ for the case of 1/1 primary resonance shown in Fig. \ref{fig_poincareivarphi}. The results (a), (b), and (c) in  Fig. \ref{fig_fourieranalysis} correspond to the lines (i), (ii), and (iii) in Fig. \ref{fig_poincareivarphi}, respectively.
As we can see in Fig. \ref{fig_poincareivarphi}, the trajectory corresponding to (i) is close to the fixed point of the $1/1$ primary resonance point, while the trajectory corresponding to (iii) lies on the boundary between the region of almost regular motion and the region of chaotic motion. 
The Fourier spectra in Fig. \ref{fig_fourieranalysis}(a), (b), and (c) exhibit clear main peaks
due to the approximately quasi-periodic nature of the almost regular motion in the island structure around the $1/1$ primary resonance.
Notably, the peak shift observed in Fig. \ref{fig_fourieranalysis}(a), (b) and (c) indicates that
the angular frequency of the main peak decreases as the distance from the fixed point increases on the Poincar\'e surface of section in Fig. \ref{fig_poincareivarphi}.

The Fourier spectrum for the trajectory corresponding to (iv) in Fig. \ref{fig_poincareivarphi} is presented in Fig.~\ref{fig_fourieranalysis}(d), which exhibits chaotic motion located outside the $1/1$ resonance region. Thus, in Fig.~\ref{fig_fourieranalysis}(d), we observe much smaller  non-vanishing frequencies proportional to $\lambda^N$, associated with the $N$-th order resonance for $N\gg 1$ than the non-vanishing frequency for the $1/1$ resonance.

Fig.~\ref{fig_fourieranalysis} illustrates that the non-vanishing angular frequencies associated with the $1/1$ resonances reach their lowest value at the threshold between the region of almost regular motion and the region of the chaotic motion. Therefore, we refer to this lowest frequency at the threshold as the minimum angular frequency. This minimum angular frequency can be defined for each resonance with an island structure on the Poincar\'e surface of section, because there is a threshold between almost regular motion and a chaotic motion 
for each resonance point\cite{Wilkie1997}. We denote the minimum angular frequency for the $M/N$ resonance as $\omega_{M,N}^{\rm min}$.  In the case of the $1/1$ primary resonance, we can roughly estimate the value of the minimum angular frequency as $\omega_{1,1}^{\rm min}=0.253$ for the system with $\lambda = 1/10$, as shown in Fig.~\ref{fig_fourieranalysis} (c).

\begin{figure*}[t]
  \begin{center}
    \includegraphics[width=10.0cm,keepaspectratio]{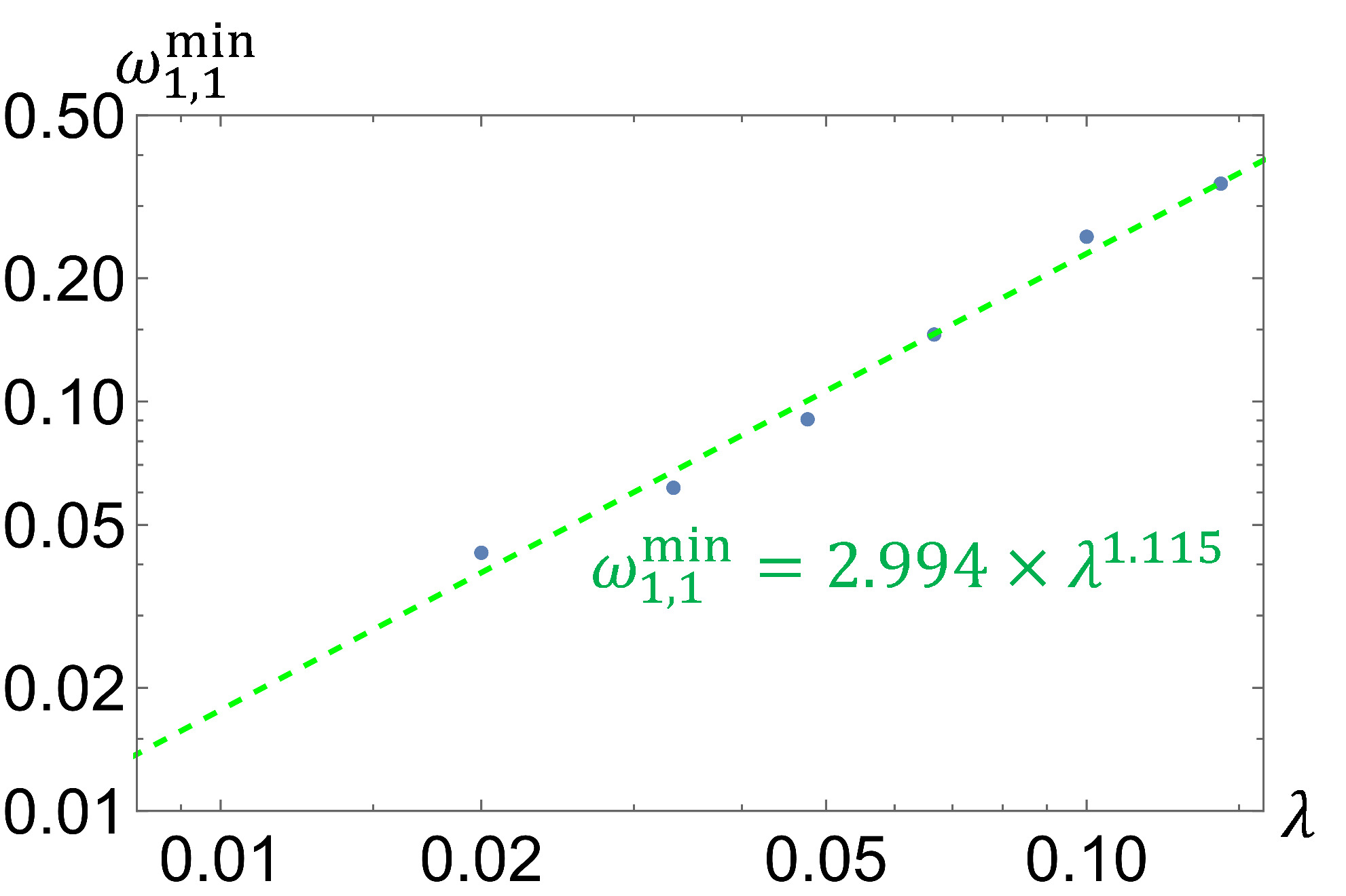}
    \caption{The dependency of $\lambda$ for the minimum angular frequency around $1/1$ primary resonance point for the dimensionless Hamiltonian $H=18.222$ and $\kappa=1/3$. The blue points are estimated values of $\omega_{1,1}^{\rm min}$ for the systems with $\lambda = 1/7, 1/10, 1/15, 1/21, 1/30$, and $1/50$. The dashed green line, given by $\omega_{1,1}^{\rm min}=2.994\times \lambda^{1.115}$, is obtained as a straight line fitted to the blue points using the least squares method. 
}
      \label{fig_eigenvaluelambda11}
  \end{center}
\end{figure*}
\begin{figure*}[ht]
  \begin{center}
    \includegraphics[width=10.0cm,keepaspectratio]{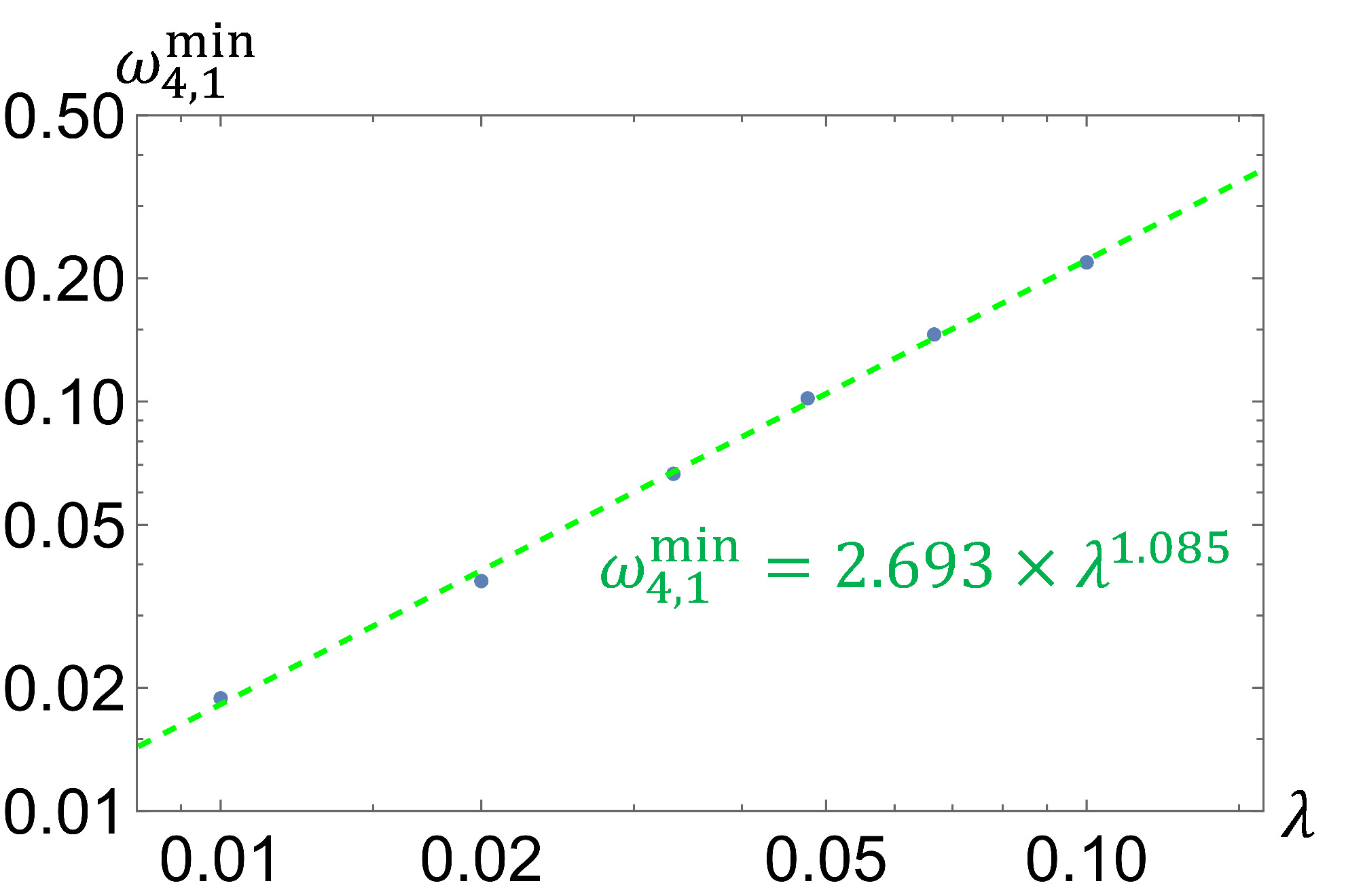}
    \caption{
The dependency of $\lambda$ for the minimum angular frequency around $4/1$ primary resonance point for the dimensionless Hamiltonian $H=18.222$ and $\kappa=1/3$. The blue points are estimated values of $\omega_{4,1}^{\rm min}$ for the systems with $\lambda = 1/10, 1/15, 1/21, 1/30, 1/50$, and $1/100$. The dashed green line, given by $\omega_{4,1}^{\rm min}=2.693\times \lambda^{1.085}$, is obtained as a straight line fitted to the blue points using the least squares method. 
}
      \label{fig_eigenvaluelambda41}
  \end{center}
\end{figure*}
\begin{figure*}[ht]
  \begin{center}
    \includegraphics[width=10.0cm,keepaspectratio]{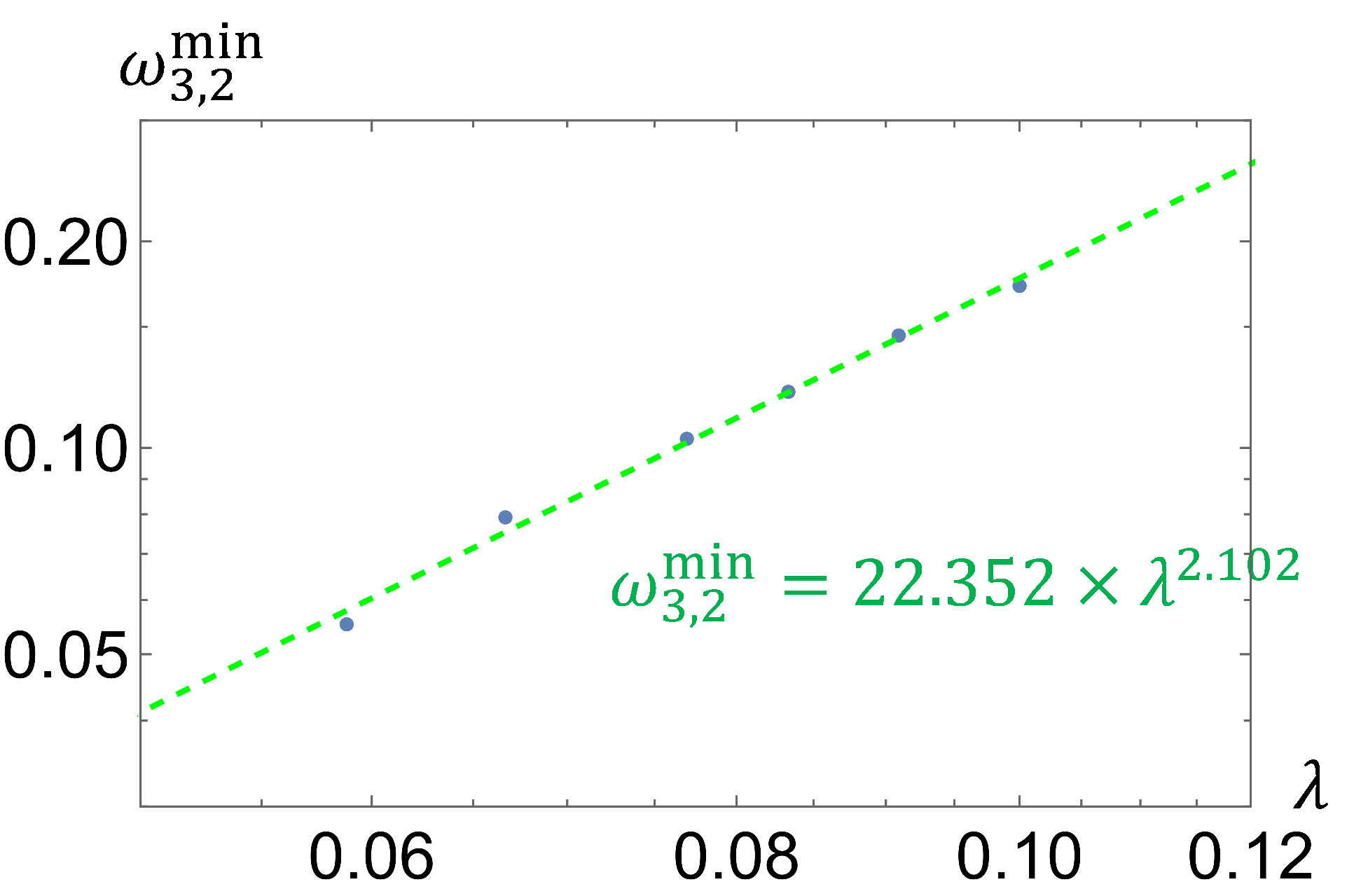}
    \caption{The dependency of $\lambda$ for the minimum angular frequency around $3/2$ resonance point for the dimensionless Hamiltonian $H=18.222$ and $\kappa=1/3$. The blue points are estimated values of $\omega_{3,2}^{\rm min}$ for the systems with $\lambda = 1/10, 1/11, 1/12, 1/13, 1/15$, and $1/17$. The dashed green line, given by $\omega_{3,2}^{\rm min}=22.352\times \lambda^{2.102}$, is obtained as a straight line fitted to the blue points using the least squares method. }
      \label{fig_eigenvaluelambda32}
  \end{center}
\end{figure*}
\begin{figure*}[ht]
  \begin{center}
    \includegraphics[width=10.0cm,keepaspectratio]{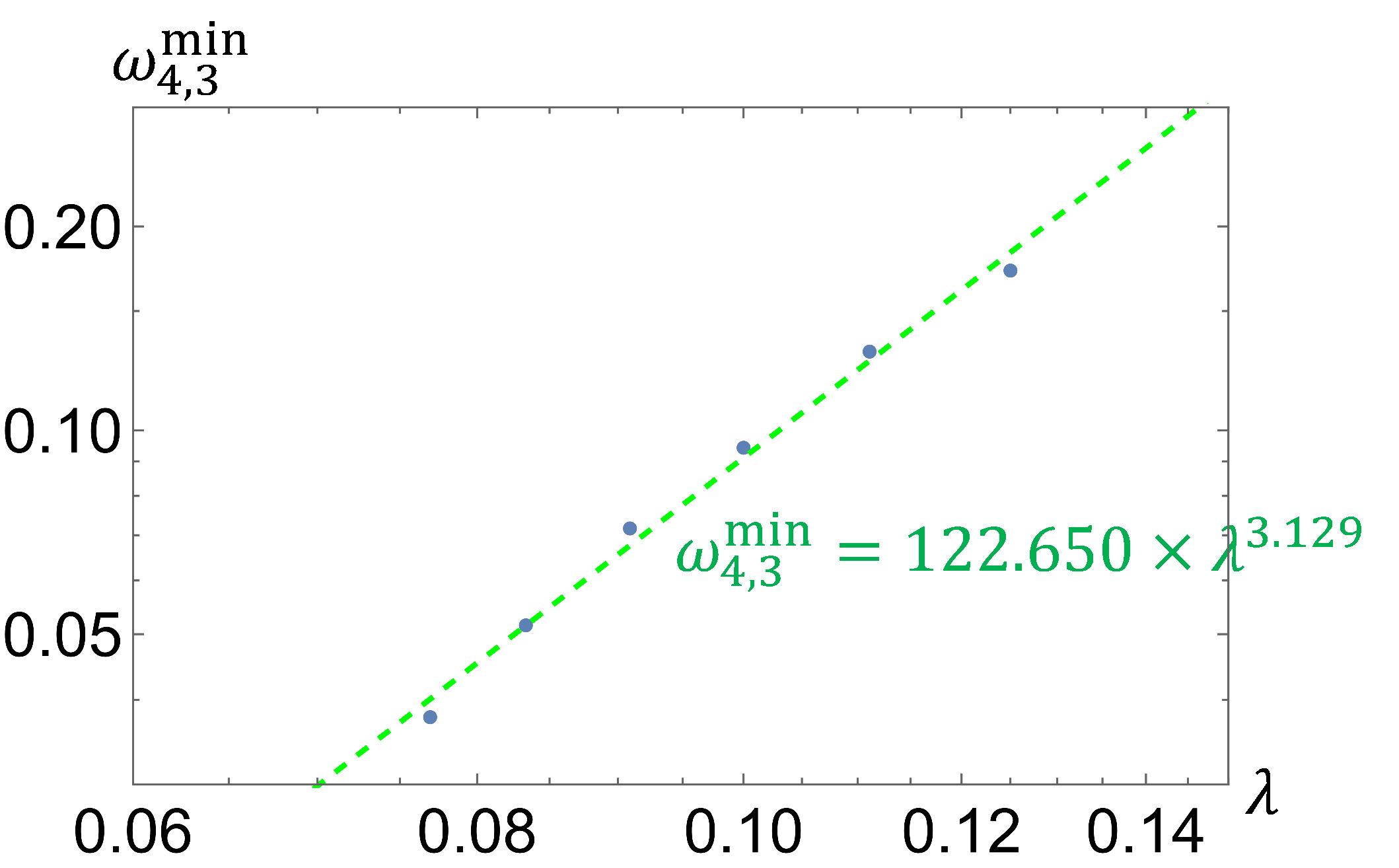}
    \caption{The dependency of $\lambda$ for the minimum angular frequency around $4/3$ resonance point for the dimensionless Hamiltonian $H=18.222$ and $\kappa=1/3$. The blue points are estimated values of $\omega_{4,3}^{\rm min}$ for the systems with $\lambda = 1/8, 1/9, 1/10, 1/11, 1/12$, and $1/13$. The dashed green line, given by $\omega_{4,3}^{\rm min}=122.650\times \lambda^{3.129}$, is obtained as a straight line fitted to the blue points using the least squares method. }
      \label{fig_eigenvaluelambda43}
  \end{center}
\end{figure*}

We have estimated the minimum angular frequencies $\omega_{1,1}^{\rm min}$ for systems with the coupling constants, $\lambda=1/7, 1/10, 1/15, 1/21, 1/30$, and $1/50$, in the case of $\kappa=1/3$ and the dimensionless Hamiltonian $H=18.222$, 
by calculating the Fourier spectra of the time evolution of trajectories for those systems. 
Fig. \ref{fig_eigenvaluelambda11} shows the $\lambda$ dependence of the minimum angular frequency around the $1/1$ primary resonance point in a double logarithmic plot. The green dashed line in Fig. \ref{fig_eigenvaluelambda11} represents a straight line fitted to those points using the least squares method. This result shows the $\lambda^b$ dependence is $b=1.115$ which is almost the same as the frequency gap for the primary resonance predicted by our theory.

In the same manner, Fig. \ref{fig_eigenvaluelambda41} shows the $\lambda$ dependence of the minimum angular frequency, $\omega_{4,1}^{\rm min}$, around the $4/1$ primary resonance point for $\lambda=1/10, 1/15, 1/21, 1/30, 1/50$, and $1/100$. This result shows the $\lambda^b$ dependence is $b=1.085$ which is again almost the same as the frequency gap for the primary resonance predicted by our theory.

In a manner similar to the case of the primary resonance, we estimate the minimum angular frequency around the $3/2$ and $4/3$ resonance points as examples of the secondary and ternary resonances. Fig. \ref{fig_eigenvaluelambda32} shows the $\lambda$ dependence of the minimum angular frequency, $\omega_{3,2}^{\rm min}$, around the $3/2$ secondary resonance point for $\lambda=1/10, 1/11, 1/12, 1/13, 1/15$, and $1/17$. This result shows the $\lambda^b$ dependence is $b=2.102$ which is almost the same as the frequency gap for the secondary resonance predicted by our theory. Fig. \ref{fig_eigenvaluelambda43} shows the $\lambda$ dependence of the minimum angular frequency, $\omega_{4,3}^{\rm min}$, around the $4/3$ ternary resonance point for $\lambda=1/8, 1/9, 1/10, 1/11, 1/12$, and $1/13$. This result shows the $\lambda^b$ dependence is $b=3.129$ which is almost the same as the frequency gap for the ternary resonance predicted by our theory.

All these results confirm our theoretical prediction of the frequency gap estimated from the eigenvalue problem of the Liouvillian.

\section{Resonance Singularity v.s. Hamiltonian}
\label{ResonanceAndHamiltonian}

We have shown the frequency gap in the spectrum of the Liouvillian for non-integrable systems. The frequency gap emerges due to the resonance singularity, which leads to the divergence of the collision operator in our analysis of the nonlinear eigenvalue problem. Eigenfunctions with zero eigenvalues in a non-perturbative integrable system vanish when a non-integrable perturbation is imposed, because the invariants of motion are destroyed by the resonant singularities.

However, there is an exception: a function of the total Hamiltonian remains the eigenfunction with the zero eigenvalue even in the presence of the non-integrable perturbation. Hence, the zero eigenvalue corresponding to this eigenfunction exists inside the frequency gap in the spectrum of the perturbed Liouvillian. It is obvious that the total Hamiltonian, as well as any functions of the total Hamiltonian, is an invariant of motion because of Eq.~(\ref{eq_ahameq}). However, it is not obvious how the total Hamiltonian remains an invariant of motion despite the presence of a resonance singularity in non-integrable systems.

Let us show the reason why the resonance singularity does not affect the total Hamiltonian from the perspective of the eigenvalue problem for the collision operator. To see the reason, we now analyze the case of the zero eigenvalue of the total Liouvillian $L_H$, i.e., $\Omega_{\alpha}^{(\bm{n})}(\bm{I}) = 0$, for a non-integrable system with $\mathcal{N}$ degrees of freedom. By using Eqs.~(\ref{eq_effl1}) and (\ref{eq_effl2}) for this case, we have the collision operator,
\begin{eqnarray}
  \psi^{(\bm{0})}(0) &=& \lambda^2 P^{(\bm{0})} L_V Q^{(\bm{0})} 
    \sum_{s=0}^{\infty}\left\{ \frac{\lambda}{-Q^{(\bm{0})}L_0 Q^{(\bm{0})}}Q^{(\bm{0})}L_V Q^{(\bm{0})} \right\}^s \nonumber \\
  & &\times \frac{1}{-Q^{(\bm{0})}L_0 Q^{(\bm{0})} } Q^{(\bm{0})}L_V P^{(\bm{0})}, \quad
   \label{eq_psi0l}
\end{eqnarray} 
where the projection operators are defined as $P^{(\bm{0})}\equiv |\bm{0}\rangle\langle \bm{0}|$ and
$Q^{(\bm{0})}\equiv \hat{I}_{\bm{\varphi}} - P^{(\bm{0})}$.
Since the denominator is zero at the resonance points, this operator generally diverges when it acts on a function in the $P^{(\bm{0})}$ space, for example, a function that depends only on the momenta $\bm{I}$.

However, we can show that when $\psi^{(\bm{0})}(0)$ acts on the Hamiltonian $H$, all the denominators in Eq. (\ref{eq_psi0l}) cancel out with the factors that appear in the numerators. This is shown in Appendix \ref{section_D}. Hence, all resonance singularities disappear  in the eigenvalue equation, and we have
\begin{eqnarray}
 \psi^{(\bm{0})}(0) P^{(\bm{0})}|H\rangle = 0.
\end{eqnarray}
As a result, the total Hamiltonian remains an invariant of motion even in the non-integrable system with the resonance singularity.

\section{Summary and Concluding Remarks}
\label{concluding}

In this paper, we have presented a method for quantitatively treating the eigenvalue problem of the Liouvillian governing the time evolution of the non-integrable pendulum. In treating the eigenvalue problem, we did not solve it directly, but introduced an auxiliary operator called the collision operator, which plays a central role in non-equilibrium statistical mechanics, and analyzed its eigenvalue problem to address the original  eigenvalue problem of the Liouvillian. The important point of our method is that we have used the nonlinearity of the eigenvalue problem, which arises from the fact that the collision operator itself depends on its eigenvalue. By focusing on this nonlinearity, we have discussed how a frequency gap appears around the zero eigenvalue of the Liouvillian as a result of the invariants of motion of the unperturbed system being destroyed by a resonant singularity caused by a non-integrable interaction. This was done without directly solving the eigenvalue problem of the collision operator, simply by exploring the conditions for the existence of the collision operator, and further quantitatively calculating the size of the gap.
Thanks to this method, we have been able to show that when the coupling constant is sufficiently small, the lower bound of the magnitude of the frequency gap in the Liouvillian spectrum at the $M/N$ resonance for any value of $N$ in our non-integrable pendulum is proportional to $\lambda^N$. Then, by numerically solving Hamilton's equations of motion,  we have confirmed that the non-vanishing minimum angular frequency in trajectories around the $M/N$ resonance point is also approximately proportional to $\lambda^N$, which is consistent with our theoretical result of the magnitude of the frequency gap.

Note that the effect of the perturbation to the eigenvalues of the unperturbed Liouvillian  is proportional to the first power of $\lambda$ for $N=1$. This result is not trivial, since after incorporating the component of  $\lambda V_0(\bm{I})$ 
in the unperturbed Hamiltonian $H_0$ as in Eq. (\ref{eq_jamvp}), it is easy to show that in the absence of degeneracy, the minimum correction by the perturbation is proportional to $\lambda^2$. In contrast, for the infinitely degenerate resonance case considered here, the resonance singularity significantly amplifies the effect of the perturbation by a factor of $1/\lambda$ compared to the case without resonance.

A similar amplification of the effect of perturbations due to degeneracy occurs in the eigenvalue problem of the Hamiltonian in quantum mechanics. However, in the case of quantum mechanics, when the degeneracy is finite, as mentioned in the introduction, one can  quantitatively evaluate the magnitude of the frequency gap around the degeneracy by solving algebraic equations of finite order. In contrast, in the case of infinitely degenerate eigenvalue problems, such as the eigenvalue problem of the Liouvillan discussed in this paper, the size of the frequency gap cannot be quantitatively evaluated by the well-known methods in quantum mechanics. We would like to emphasize that we have been able to obtain the size by utilizing the nonlinearity of the eigenvalue problem of the collision operator as shown here, rather than by using the well-known methods in quantum mechanics.

It should be noted that the above conclusion is obtained for a specific model where the Fourier series of the interaction, $V_{l_1, l_2}(I_1, I_2)$, with respect to $\varphi_2$ has only Fourier exponents with $l_2= \pm1$. In a more general case, the above results need to be modified depending on the Fourier series of the interaction with respect to $\varphi_2$. However, it is clear that the modifications can be obtained by a straightforward extension of the above arguments. We would also like to mention that our treatment presented here is not limited to two-degree-of-freedom systems but can also be extended to more general systems with more than two degrees of freedom through a straightforward extension of the above arguments.

Regarding the collision operator used here, we would like to mention a crucial difference between its use in non-equilibrium statistical mechanics for systems with infinite degrees of freedom, known as the thermodynamic limit, and its use in non-integrable systems with finite degrees of freedom, as discussed in this paper. The biggest difference is that, as stated between Eq. \eqref{eq_ef0} and Eq. \eqref{eq_ev00}, $\Delta_j(I_j)$, which represents the discreteness of the spectrum of the unperturbed Liouvillian, is finite in our paper, whereas in the thermodynamic limit, $\Delta_j(I_j)$ becomes infinitesimal, and in non-equilibrium statistical mechanics we deal with Fourier integrals with respect to continuous spectra rather than Fourier series as discussed in this paper. In the case of non-equilibrium statistical mechanics with a continuous spectrum, division by zero in the frequency denominator at the resonance singularity appearing in the propagator in the solution of the Liouville equation or in the eigenvalue problem of the Liouvillian can be treated in a mathematically well-defined form by analytical continuation using the Cauchy integral derived from the Fourier integral \cite{Petrosky1996,Balescu1963}. This point is crucially different from non-integrable systems with a finite degree of freedom as discussed here, where division by zero in the Fourier series makes the Fourier series mathematically meaningless. And in the case of non-equilibrium statistical mechanics with a continuous spectrum, it can be shown that the time-symmetry is broken due to the resonance singularity by the difference in the direction of analytical continuation of the Cauchy integral on the complex frequency plane. On the other hand, in the non-integrable systems with a finite degree of freedom as considered here, the divergence of the Fourier series caused by the resonance singularity is the mechanism by which the invariants of motion are destroyed in a non-integrable system. Therefore, we have named the system with a continuous spectrum, which was taken to have a thermodynamic limit, {\it the large Poincar\'e systems}, to distinguish it from the non-integrable systems with  a finite degree of freedom as discussed here (see Ref.\cite{Petrosky1996}for more details).

Note that in the systems considered here, the size of the frequency gap around the zero eigenvalue at the $M/N$ resonance is proportional to $\lambda^N$, and as a result, the resonance effects become smaller as $N$ becomes larger. On the other hand, we can see that as $N$ increases, the resonance effect diminishes, as seen from the size of the islands on the Poincar\'e surfaces of section in Figs.~\ref{fig_poincareivarphi} and \ref{fig_poincareivarphi40}.  In fact, from numerical calculations, it is easy to verify that the width in the $I_1$ direction of the island around the $M/N$ resonance is of the order of $\lambda^{N/2}$. This suggests that the size of the frequency gap appearing in the eigenvalue problem of the Liouvillian is also related to the size of the island on the Poincar\'e surface of section in trajectory dynamics. We would like to discuss a detailed analytical research into this interesting relationship between these two as a future topic.

Finally, let us make a comment on the relation of our analysis of the invariants of motion to the KAM theory\cite{Barrar1970}.  In our analysis based on the Liouvillian that consists of the differential operator of the generalized momenta, we are investigating the mechanism behind the destruction of invariants of motion that are continuous functions of the generalized momenta. In other words, we are not investigating the possibility of the existence of invariant tori that are not continuous functions of the generalized momenta discussed in the KAM theory\cite{Barrar1970}. Some aspects of the relationship between invariant tori in the KAM theory and the collisional invariant have been discussed in the Ref. \cite{CollisionalInvariant1982}. We hope that we can discuss the relation between the two in more detail in the future.

\section*{Acknowledgments}

We thank Professor S. Tanaka, Professor K. Kanki and Professor S. Garmon for many discussions and for their fruitful suggestions. This work was supported by JST, the establishment of university fellowships towards the creation of science technology innovation, Grant Number JPMJFS2138.

\appendix

\section{Canonical Transformation}
\label{section_B}

We explain the canonical transformation in Section \ref{Nonintpendulum}, dividing it into four steps.

First, we canonically transform from the variables $(X, P_X)$ to the action-angle variables $(\phi, J_{\phi})$ in terms of the harmonic oscillator as follows,
\begin{eqnarray}
  X &=& \sqrt{ \frac{2J_{\phi}}{m\omega_h} } \cos \phi ,
    \label{eq_transxphi} \\
  P_X &=& -\sqrt{ 2m\omega_h J_{\phi} } \sin \phi .
    \label{eq_transpxjphi}
\end{eqnarray}
The energy of the unperturbed harmonic oscillator is ${H}_2=\omega_h J_{\phi}$ and the parameter $\omega_h$ is the angular frequency of the unperturbed harmonic oscillator.

Second, we transform the Hamiltonian, the time, and each variable into dimensionless quantities by
\begin{eqnarray}
  \tilde{H} &=& {H}_t / {H}_{{\rm sx}} ,
    \label{eq_transham} \\
  \tau &=& \omega_0 t ,
    \label{eq_transtime} \\
  \Theta' &=& \Theta ,
    \label{eq_transtheta} \\
  Y' &=& \omega_0 P_{\Theta} / {H}_{{\rm sx}} ,
    \label{eq_transy1ptheta} \\
  \theta_2 &=& \phi ,
    \label{eq_transalpha2phi} \\
  J_2 &=& \omega_0 J_{\phi} / {H}_{{\rm sx}} ,
    \label{eq_transy2jphi}
\end{eqnarray}
where, ${H}_{\rm sx}=2ml^2\omega_0^2$ is the energy of the unperturbed pendulum at the separatrix. The energy of the unperturbed pendulum is $\tilde{H}_1= Y'^2+(1-\cos \Theta')/2$. Note that the variables $(\theta_2, J_2)$ are action-angle variables with respect to the dimensionless time $\tau$. The energy of the unperturbed harmonic oscillator is $\tilde{H}_2= J_2 / \kappa$.

Next, we introduce the action-angle variables of the pendulum $(\theta_1, J_1)$ with respect to the dimensionless time $\tau$. We define the action variable of the pendulum as
\begin{eqnarray}
  J_1 \equiv \frac{1}{2\pi} \oint Y'(H_1, \Theta') d\Theta' ,
    \label{eq_j1def}
\end{eqnarray}
where
\begin{eqnarray}
  Y'(H_1, \Theta') = \pm \frac{1}{c} \sqrt{1-c^2\sin^2 \frac{\Theta'}{2}} .
    \label{eq_yppm}
\end{eqnarray}
Calculating the integration in Eq. (\ref{eq_j1def}), we can derive
\begin{eqnarray}
  J_1=
    \begin{cases}
      \frac{2}{\pi c}E(c) & c \le 1 \\
      \frac{4}{\pi} \{ E(1/c) - (1-\frac{1}{c^2}) K(1/c) \} & c>1,
    \end{cases}
    \label{eq_j1cal}
\end{eqnarray}
where $E(c)$ is the complete elliptic integral of the second kind. Because of $c=1/\sqrt{\tilde{H}_1}$, we only obtain the representation of the implicit function in terms of the Hamiltonian of the pendulum with respect to the action variable.

Introducing the generating function,
\begin{eqnarray}
  &&\hspace{-15pt}W_1(\Theta', J_1) \nonumber \\
    &=& \int Y' d\Theta' \nonumber \\
    &=&
      \begin{cases}
        \pm \frac{2}{c(J_1)}E(\Theta'/2, c(J_1)) & c \le 1 \\
        \pm 2\left \{ E(\Theta'/2, 1/c(J_1)) \right. \\ 
          \left. \hspace{20pt} - \left(1-\frac{1}{c(J_1)^2}\right)F(\Theta'/2, 1/c(J_1)) \right \} \\
          \hspace{20pt} + \theta_{1,0} J_1& c>1,
      \end{cases}  \nonumber \\
      \label{eq_gen1c} \\
  \theta_{1,0}
    &=&
      \begin{cases}
        0 & P_{\Theta}\ge 0 \\
        \pi & P_{\Theta}<0 ,
      \end{cases} 
\end{eqnarray}
we obtain the canonical transformations from $(\Theta', Y')$ to the action-angle variables $(\theta_1, J_1)$ (in the case of $c\le1$) as
\begin{eqnarray}
  \theta_1 &=& \pm \frac{\pi}{K(c)}F(\Theta'/2, c) ,
    \label{eq_theta11} \\
  \sin \frac{\Theta'}{2} &=& \pm {\rm sn}\left(\frac{K(c)}{\pi} \theta_1, c\right) ,
    \label{eq_sinthetap1} \\
  Y' &=& \pm \frac{1}{c} {\rm dn}\left(\frac{K(c)}{\pi} \theta_1, c\right),
    \label{eq_yp1}
\end{eqnarray}
and (in the case of $c>1$) as
\begin{eqnarray}
  \theta_1 &=& \pm \frac{\pi}{2K(1/c)}F(\sin^{-1} (c\sin(\Theta'/2)), 1/c) + \theta_{1,0}, \hspace{20pt}
    \label{eq_theta12} \\
  \sin \frac{\Theta'}{2} &=& \pm \frac{1}{c} {\rm sn}\left(\frac{2K(1/c)}{\pi}(\theta_1-\theta_{1,0}), 1/c\right) ,
    \label{eq_sinthetap2} \\
  Y' &=& \pm \frac{1}{c} {\rm cn}\left(\frac{2K(1/c)}{\pi}(\theta_1-\theta_{1,0}), 1/c\right), 
    \label{eq_yp2}
\end{eqnarray}
where $F(\varphi, c)$ and $E(\varphi, c)$ are the elliptic integral of the first kind and the second kind, respectively.  Here, Eqs. (\ref{eq_theta11}) and (\ref{eq_theta12}) takes a positive sign when $P_{\Theta} \ge 0$ or $Y'\ge0$ and it takes a negative sign when $P_{\Theta} < 0$ or $Y'<0$. 

In the case of the libration of the pendulum ($c>1$), we introduce the constant $\theta_{1,0}$ in order to take $\theta_1=\pi/2$ when $P_{\Theta} = 0$ or $Y'=0$, i.e., $\theta_{1,0}$ ensures the continuity of $\theta_1$ at $\theta_1=\pi/2$. Eq. (\ref{eq_theta11}) determines the range of $\theta_1$ in case of the rotational motion ($c \le 1$) as $-\pi \le \theta_1 \le \pi$, while Eq. (\ref{eq_theta12}) determines the range of $\theta_1$ in case of the librational motion ($c > 1$) as $-\pi/2 \le \theta_1 \le 3\pi/2$. This constant $\theta_{1,0}$ allows for the range of the angle $\theta_1$ as $2\pi$ in case of the libration of the pendulum.

Finally, by transforming the action-angle variables $(J_1, J_2, \theta_1, \theta_2)$ to the variables $(I_1, I_2, \varphi_1, \varphi_2)$ using the generating function,
\begin{eqnarray}
  & &\hspace{-15pt} W_2(\theta_1, \theta_2, I_1, I_2)  \nonumber \\
  & &=\begin{cases}
      \frac{2E(c)}{\pi} I_1 \theta_1 + I_2 \theta_2 & c \le 1 \\
      \frac{4}{\pi} \left\{ E(1/c)-\left(1-\frac{1}{c^2}\right)K(1/c) \right\} \theta_1 + I_2 \theta_2 & c>1,
    \end{cases} \nonumber \\
\end{eqnarray}
we can obtain the following canonical transformations,
\begin{eqnarray}
  \varphi_1 &=& 
    \begin{cases}
      \frac{2K(c(J_1))}{\pi}\theta_1 & c \le 1 \\
      \frac{4K(1/c(J_1))}{\pi c(J_1)}\theta_1 & c>1,
    \end{cases}
    \label{eq_transalpha1theta1} \\
  I_1 &=& 1/c(J_1)  ,
    \label{eq_transy1j1} \\
  \varphi_2 &=& \theta_2 ,
    \label{eq_transalpha2theta2} \\
  I_2 &=& J_2 ,
    \label{eq_transy2j2} 
\end{eqnarray}
and the relation,
\begin{eqnarray}
  \varphi_{1,0}(J_1)&=& \frac{4K(1/c(J_1))}{\pi c(J_1)}\theta_{1,0} .
\end{eqnarray}

As mentioned before, since $\theta_1$ and $\theta_2$ are angle variables, we have the periodicity of the angles $\varphi_j$ ($j=1, 2$), $2\pi / \Delta_j(I_j)$, as
\begin{eqnarray}
  \Delta_1(I_1)&=&
    \begin{cases}
      \frac{\pi}{2K(c)} & c \le 1 \\
      \frac{\pi c}{4K(1/c)} & c>1,
    \end{cases}
    \label{eq_delta1c} \\
  \Delta_2 &=& 1
    \label{eq_delta2c}
\end{eqnarray}
from Eqs. (\ref{eq_transalpha1theta1}) and (\ref{eq_transalpha2theta2}).

\section{The Canonical Variables $(I_j, \varphi_j)$ v.s. Action-Angle Variables $(J_j, \theta_j)$}
\label{section_a}

In the case where the discreteness of the spectrum $\Delta_j(I_j)$ depends on $I_j$, our canonical set of the variables $(\bm{I}, \bm{\varphi}) $ is not a set of the action-angle variables $(\bm{J}, \bm{\theta})$. In this appendix, we will show that, if $\Delta_j(I_j)$ is finite, it is always possible to perform a canonical transformation from our canonical set of the variables to the action-angle variables.

Indeed, let us consider the following generating function of the canonical transformation
\begin{eqnarray}
  W(\bm{I}, \bm{\theta}) = -\sum_{j=1}^\mathcal{N} \int   dI_j\frac{1}{\Delta_j(I_j)} \theta_j .
    \label{eq_genfunca0} 
\end{eqnarray}
Then, we have
\begin{eqnarray}
  \varphi_j =  - \frac{\partial W}{\partial I_j} = \theta_j /\Delta_j(I_j)
    \label{eq_genfunca01}
\end{eqnarray}
and
\begin{eqnarray}
  J_j =  - \frac{\partial W}{\partial \theta_j} = \int  dI_j \frac{1}{\Delta_j(I_j)} 
    \label{eq_genfunca02}
\end{eqnarray}
or
\begin{eqnarray}
  \frac{d I_j}{d J_j} = \Delta_j(I_j) .
    \label{eq_genfunca03}
\end{eqnarray}
For the action-angle variables, the periodicity with respect to the angle variable is fixed to a constant $2\pi$. Hence, if we use the action-angle variables in the representation of the Liouvillian, the discreteness of the spectrum is equal to $1$ so that it does not depend of the action variable. As a result, we have the following matrix element for the perturbed Liouvillian corresponding to Eq. (\ref{eq_matrixelementprime1}),
\begin{eqnarray}
  &&\hspace{-20pt} \Braket{\bm{n} \hspace{-1pt} \mid L_V \mid \hspace{-1pt} \bm{n'} } \nonumber \\
  &=&-\left( V_{\bm{n}-\bm{n'}}(\bm{J}) \left(\bm{n}-\bm{n'} \right) \cdot \frac{\partial}{\partial \bm{J}} - \bm{n'} \cdot \frac{\partial V_{\bm{n}-\bm{n'}}(\bm{J})}{\partial \bm{J}} \right) . \nonumber \\
    \label{eq_matrixelementactionangle}
\end{eqnarray}

As will be shown in Appendix \ref{section_D}, this representation is useful for demonstrating why the total Hamiltonian can remain an invariant of motion despite the fact that other invariants of motion are broken by the resonance singularity.

Nevertheless, our original variables $(\bm{I}, \bm{\varphi})$ may have an advantage compared to the action-angle variables. 
 For example, as shown in Appendix \ref{section_B}, the explicit form of the Hamiltonian of the unperturbed pendulum can be written in terms of our variables, while in terms of the action-angle variables, we only obtain a representation of an implicit function in terms of the Hamiltonian. Furthermore, we can define the value of $I_1$ at the separatrix between the rotational motion and the librational motion of the pendulum as $I_1=1/c=1$, while the action variable of the pendulum cannot be defined at the separatrix.

\section{The Convergence Condition of the Coefficient of the Collision Operator}
\label{section_C}

In this Appendix, we provide a proof of Eq. \eqref{DMNj}. By the straightforward extension of Eq. \eqref{colli42}, we can see that the denominator in the intermediate state 
$Q^{(\bm{n})}|\bm{n} +\bm{l} + \bm{l'} +\cdots + \bm{l}^{[k]} \rangle$, which is reached after $k$ interactions with $1 \le k < N$ 
in the transition starting with $|\bm{n} \rangle $ located at the left-hand side of its corresponding expression in Eq. \eqref{colli42}, is proportional to
\begin{eqnarray}
  & &C(I_{M,N}, I_2) \nonumber \\
  & &\hspace{5pt} - (n_1 + l_1 + l'_1 +\cdots + l_1^{[k]} ) N  - (n_2 + l_2+ l'_2 +\cdots + l_2^{[k]} ) M \nonumber \\
  & & \qquad = C(I_{M,N}, I_2) - p N  -  qM,   
   \label{IntermediateDenominator}
\end{eqnarray}
where
\begin{eqnarray}
  p 
   &\equiv& l_1 + l'_1 +\cdots + l_1^{[k]}, \nonumber \\
  q 
   &\equiv& l_2 + l'_2 +\cdots + l_2^{[k]},  
 \label{pq}
\end{eqnarray}
and we have used the relation $n_1 N + n_2 M =0$. Furthermore, if we recall that the Fourier coefficient of the interaction, $V_{l_1, l_2}(I_1, I_2)$, with respect to $\varphi_2$ has only the Fourier index with $l_2= \pm1$ in our non-integrable pendulum, we have an integer $q$ that satisfies $0 \le |q|  \le k < N$. Hence, if the following equality satisfies,
\begin{eqnarray}
  pN+qM = 0,
 \label{pqresonance}
\end{eqnarray}
i.e., $q/p =-N/M$, then the denominator including the factor \eqref{IntermediateDenominator} reduces to the minimum value $\omega_2 C(I_{M,N}, I_2)/M$ in the corresponding expression in Eq. \eqref{colli42}.  However, it is impossible, because there is no integer $q$ with $|q|< N$, since $M$ and $N$ are coprime integers and $p=q=0$ is excluded because of the condition $Q^{(\bm{n})}$ in the intermediate states. While, if $q=\pm N$, then Eq. \eqref{pqresonance} holds in case of $p=\mp M$. Hence, this proves that we need at least $N$ times transitions  of $\lambda L_V$, from the state $| \bm{n} \rangle $ to the intermediate state associated with the $M/N$ resonance with any value of $N$. As a result, Eq. \eqref{DMNj} holds for any value of $N$.

\section{Peculiaritiy of Hamiltonians in Resonance Singularities}
\label{section_D}

The proof that the total Hamiltonian $H$ is an invariant of motion despite being a non-integrable system is obvious from the following relation (see Eqs. (\ref{eq_cl}) and (\ref{eq_ahameq})),
\begin{equation}
  i\frac{d}{d t} H(\bm{I}, \bm{\varphi}) = -L_H H(\bm{I}, \bm{\varphi})  = 0.
\end{equation}
However, it is not obvious why $H$ can be an invariant of motion without being affected by the resonance singularity, even though it is a non-integrable system. In this Appendix, we clarify why the resonance singularity does not affect the total Hamiltonian from the perspective of the eigenvalue problem for the collision operator introduced in this paper.

If a physical quantity $A(\bm{I}, \bm{\varphi})$ is an invariant of motion, then $L_H A(\bm{I}, \bm{\varphi})=0$. In the unperturbed system, any function that depends only on the momenta is an invariant of motion. Thus, the eigenfunction represented by the function of the momenta belongs to the $P^{(\bm{0})}\equiv | \bm{0} \rangle \langle \bm{0} |$ subspace introduced in Eq. (\ref{eq_projectionpn}). Because of $L_H H=0$ from the definition of the Liouvillian (\ref{eq_cl}), the following relation should hold even for non-integrable systems with the resonance singularity,
\begin{equation}
  \psi^{(\bm{0})}(0) P^{(\bm{0})} \mid \hspace{-3pt}  H(\bm{I}) \rangle = \psi^{(\bm{0})}(0) \mid \hspace{-3pt} H_0(\bm{I}) \rangle = 0 ,
    \label{eq_psiham0}
\end{equation}
where $\mid \hspace{-3pt} H_0(\bm{I}) \rangle = P^{(\bm{0})} \mid \hspace{-3pt} H(\bm{I}) \rangle$, and
the bra-ket notation of the total Hamiltonian and the unperturbed Hamiltonian are given by
\begin{eqnarray}
  \mid \hspace{-3pt} H(\bm{I}) \rangle &=& \mid \hspace{-3pt} H_0(\bm{I}) \rangle + \lambda \sqrt{\frac{(2\pi)^{\mathcal{N}}}{\prod_{j=1}^\mathcal{N} \Delta_j(I_j)}} {\sum_{\bm{n}}}^{'} V_{\bm{n}}(\bm{I}) \mid \hspace{-3pt} \bm{n} \rangle, 
    \label{eq_hamiltonianbraket}
\end{eqnarray}
with
\begin{eqnarray}
   \mid \hspace{-3pt} H_0(\bm{I}) \rangle &=& \sqrt{\frac{(2\pi)^{\mathcal{N}}}{\prod_{j=1}^\mathcal{N} \Delta_j(I_j)}} H_0(\bm{I}) \mid \hspace{-3pt} \bm{0} \rangle.  
    \label{eq_hamiltonian0braket}
\end{eqnarray}

In the following discussion, we will show that relation (\ref{eq_psiham0}) holds in a well-defined form, as a result of which all resonance singularities arising from the small denominators in the series expansion of equation (\ref{eq_psiham0}) disappear. Furthermore, we will show that the elimination of the resonance singularity occurs only in the case of $H$ and its functions, while for other functions of the momenta, the resonance singularity leads to divergence.

As can be seen from the discussion in the paragraph below Eq. (\ref{eq_nreduction}), the eigenfunctions of the unperturbed Liouvillian are Fourier basis for the Fourier transformation, so the integer vector $\mid \hspace{-3pt} \bm{n} \rangle$ is the same in both the $(\bm{J},\bm{\theta})$-representation and the $(\bm{I},\bm{\varphi})$-representation. Moreover, as shown in Eq. (\ref{eq_matrixelementactionangle}), we can represent the matrix element of the perturbed Liouvillian more simply in the $(\bm{J},\bm{\theta})$-representation than in the $(\bm{I},\bm{\varphi})$-representation. Hence, we will use the action-angle variables in the $(\bm{J}, \bm{\theta})$-representation instead of the $(\bm{I},\bm{\varphi})$-representation in the following discussion.

Acting the collision operator (\ref{eq_psi0l}) on $\mid \hspace{-3pt} H_0(\bm{J}) \rangle$, we have
\begin{eqnarray}
  \psi^{(\bm{0})}(0) \mid \hspace{-3pt} H_0(\bm{J}) \rangle
    &=&\lambda^2 P^{(\bm{0})} L_V Q^{(\bm{0})} \frac{1}{- Q^{(\bm{0})} L_0 Q^{(\bm{0})}} \nonumber \\
    & &\times \sum _{s=0}^\infty \left( \lambda Q^{(\bm{0})} L_V Q^{(\bm{0})} \frac{1}{- Q^{(\bm{0})} L_0 Q^{(\bm{0})}} \right)^s \nonumber \\
    & & \hspace{20pt} \times Q^{(\bm{0})} L_V P^{(\bm{0})} \mid \hspace{-3pt} H_0(\bm{J}) \rangle .
    \label{eq_effl1000}
\end{eqnarray}

Using the matrix element of the perturbed term of the Liouvillian in terms of the action-angle variables in Eq. (\ref{eq_matrixelementactionangle}), we calculate the lowest order of $\lambda$ in Eq. (\ref{eq_effl1000}) as
\begin{eqnarray}
  & &\hspace{-15pt}\lambda^2 \psi_2^{(\bm{0})}(0) \mid \hspace{-3pt} H_0(\bm{J}) \rangle \nonumber \\
    &=& \lambda^2 P^{(\bm{0})} L_V Q^{(\bm{0})} \frac{1}{- Q^{(\bm{0})} L_0 Q^{(\bm{0})}} Q^{(\bm{0})} L_V P^{(\bm{0})} \mid \hspace{-3pt} H_0(\bm{J}) \rangle \nonumber \\
    &=&-\lambda^2 (2\pi)^{\frac{\mathcal{N}}{2}} {\sum_{\bm{n}}}^{'} \mid \bm{0} \rangle \braket{\bm{0} \hspace{-3pt} \mid L_V \hspace{-1pt} \mid \hspace{-1pt}  \bm{n}} \frac{1}{w^{(\bm{n})}(\bm{J})} \braket{\bm{n} \hspace{-1pt} \mid L_V \hspace{-1pt} \mid \hspace{-1pt}  \bm{0}} H_0(\bm{J}) \nonumber \\
    &=&\lambda^2 (2\pi)^{\frac{\mathcal{N}}{2}} {\sum_{\bm{n}}}^{'} \mid \bm{0} \rangle \braket{\bm{0} \hspace{-3pt} \mid L_V \hspace{-1pt} \mid \hspace{-1pt}  \bm{n}} V_{\bm{n}}(\bm{J}) \frac{1}{w^{(\bm{n})}(\bm{J})} \bm{n} \cdot \frac{\partial H_0(\bm{J})}{\partial \bm{J}} , \nonumber \\
      \label{eq_effl10020}
\end{eqnarray}
where the unperturbed Hamiltonian vector in the $(\bm{J},\bm{\theta})$-representation is given as
\begin{eqnarray}
   \mid \hspace{-3pt} H_0(\bm{J}) \rangle &=& (2\pi)^{\frac{\mathcal{N}}{2}} H_0(\bm{J}) \mid \hspace{-3pt} \bm{0} \rangle ,
    \label{eq_hamiltonian0braket_jtheta}
\end{eqnarray}
and we use the relation between the unperturbed Hamiltonian and that of the bra-ket notation:
\begin{eqnarray}
  H_0(\bm{J}) = \langle \bm{0}| H_0(\bm{J})\rangle .
\end{eqnarray}
Here, as explained in Section \ref{liouville}, the prime sign in ${\sum}^{'}$ in Eq. (\ref{eq_effl10020}) means to exclude the projection with $\bm{n} = \bm{0}$. Since the unperturbed eigenvalue in terms of the action-angle variables is $w^{(\bm{n})}(\bm{J})=\bm{n} \cdot \bm{\omega}(\bm{J})=\bm{n} \cdot \partial H_0(\bm{J}) / \partial \bm{J}$, the frequency denominator and its numerator in Eq. (\ref{eq_effl10020}) are canceled. Thus, the effect of the resonance singularity vanishes in case of the Hamiltonian $\mid \hspace{-3pt} H_0(\bm{J}) \rangle$. Therefore, we can obtain
\begin{eqnarray}
  \lambda^2 \psi_2^{(\bm{0})}(0) \mid \hspace{-3pt} H_0(\bm{J}) \rangle
    &=&\lambda^2 (2\pi)^{\frac{\mathcal{N}}{2}} {\sum_{\bm{n}}}^{'} \mid \bm{0} \rangle \braket{\bm{0} \hspace{-3pt} \mid L_V \hspace{-1pt} \mid \hspace{-1pt}  \bm{n}} V_{\bm{n}}(\bm{J}) . \nonumber \\
      \label{eq_effl10021}
\end{eqnarray}
We also calculate Eq. (\ref{eq_effl10021}) as
\begin{eqnarray}
  & &\hspace{-25pt} \lambda^2 \psi_2^{(\bm{0})}(0) \mid \hspace{-3pt} H_0(\bm{J}) \rangle \nonumber \\
    &=&\lambda^2 (2\pi)^{\frac{\mathcal{N}}{2}} {\sum_{\bm{n}}}^{'} \mid \bm{0} \rangle \Big\{ \bm{n} \cdot \frac{\partial}{\partial \bm{J}} \left( V_{-\bm{n}}(\bm{J}) V_{\bm{n}}(\bm{J}) \right) \Big\} . \nonumber \\
    \label{eq_effl1002}
\end{eqnarray}
Since $\bm{n} \cdot \frac{\partial}{\partial \bm{J}} \left( V_{-\bm{n}}(\bm{J}) V_{\bm{n}}(\bm{J}) \right)$ in Eq. (\ref{eq_effl1002}) has the symmetry for $\bm{n}$,
\begin{eqnarray}
\bm{n} \cdot \frac{\partial}{\partial \bm{J}} \left( V_{-\bm{n}}(\bm{J}) V_{\bm{n}}(\bm{J}) \right)
  =-(-\bm{n}) \cdot \frac{\partial}{\partial \bm{J}} \left( V_{\bm{n}}(\bm{J}) V_{-\bm{n}}(\bm{J}) \right) , \nonumber \\
  \label{eq_effl100symmetry}
\end{eqnarray}
we can obtain 
\begin{eqnarray}
  \lambda^2 \psi_2^{(\bm{0})}(0) \mid \hspace{-3pt} H_0(\bm{J}) \rangle =0 .
\end{eqnarray}
By performing the same calculations for higher-order terms of $\lambda$, we can derive
\begin{eqnarray}
  \psi^{(\bm{0})}(0) \mid \hspace{-3pt} H_0(\bm{J}) \rangle = 0 .
\end{eqnarray}
Note that in case of the action variable $J_j$, which is an invariant of motion for the unperturbed system, the unperturbed eigenvalue $w^{(\bm{n})}(\bm{J})$ in its denominator in $\psi^{(\bm{0})}(0) \mid \hspace{-3pt} J_j \rangle$ instead of Eq. (\ref{eq_effl10020}) does not vanish so that this expression diverges due to the resonance singularity.

Therefore, the total Hamiltonian remains an invariant of motion regardless of the non-integrability caused by the perturbation, while the momenta $\bm{J}$ are no longer invariants of motion. Thus, the perturbation breaks all invariants of motion except the total Hamiltonian due to the resonance effect.


\begin{thebibliography}{99}
\bibitem{Petrosky2004}{For the case we consider the ensemble average of a physical quantity over the probability density $\rho({\bm I},{\bm \varphi},t)$ in the Liouvillian formalism, we take the integration over the generalized momentum ${\bm I}$ which is a continuous variable.  For this case,  the resonance singularity appearing in the denominator of the propagator in the solution of the Liouville equation, as well as in the eigenvalue problem of the Liouvillian, has completely different mathematical structure form the case of the discrete spectrum discussed in this paper. Indeed, for the case of the integration over the continuous variable the denominator leads to distributions, i.e., a generalized functions, such as the $\delta$ function and the principal part with respect to the frequency, which are associated with the Cauchy integral over the generalized momentum.  Therefore, we have to teat the effect of  the resonance singularity in a way that is different from the treatment presented in this paper.  See more detail in Ref. \cite{ChynBiu2004}.
}
\bibitem{collision11}{Even though the eigenvalue equation of the collision operator is nonlinear with respect to the eigenvalue, there is only one eigenvalue for each eigenfunction for the following two reasons: (i) The eigenvalue problem for the original Liouvillian $L_H$ has only one eigenvalue for each eigenfunction because the Liouvillian is a linear operator.
(ii) The eigenfunction $P^{(\bm{n})}|\Phi_{\alpha}^{(\bm{n})} \rangle$ of the collision operator is the $P^{(\bm{n})}$ component of the Liouvillian eigenfunction  $|\Phi_{\alpha}^{(\bm{n})} \rangle$ with the same eigenvalue.}
\end{thebibliography}

\end{document}